\documentclass[twocolumn,superscriptaddress,a4paper]{revtex4-1} % preprint  twocolumn
\usepackage{graphicx}
\usepackage{subfigure}
\usepackage{color}
\usepackage{amsmath}
\usepackage{bm}
\usepackage{setspace}
\usepackage{bibunits}
\usepackage[hypertexnames=false]{hyperref}

\begin{document}

\title{Particle correlations and evidence for dark state condensation in a cold dipolar exciton fluid}
\author{Yehiel Shilo}
\author{Kobi Cohen}
\author{Boris Laikhtman}
\affiliation{Racah Institute of Physics, The Hebrew University of Jerusalem, Jerusalem 91904, Israel}
\author{Ronen Rapaport}
\affiliation{Racah Institute of Physics, The Hebrew University of Jerusalem, Jerusalem 91904, Israel}
\affiliation{Applied Physics Department, The Hebrew University of Jerusalem, Jerusalem 91904, Israel}
\author{Ken West}
\author{Loren Pfeiffer}
\affiliation{Department of Electrical Engineering, Princeton University, Princeton, New Jersey 08544, USA}

\begin{abstract}
In this paper we show experimental evidence of a few correlation regimes of a cold dipolar exciton fluid, created optically in a semiconductor bilayer heterostructure. In the higher temperature regime, the average interaction energy between the particles shows a surprising temperature dependence which is an evidence for correlations beyond the mean field model. At a lower temperature, there is a sharp increase in the interaction energy of optically active excitons, accompanied by a strong reduction in their apparent population. This is an evidence for a sharp macroscopic transition to a dark state as was suggested theoretically.
\end{abstract}

\maketitle
\begin{bibunit}
Different collective many-body effects in Bose quantum fluids of atoms \cite{Bloch08} and exciton-polaritons \cite{Deng10} have been observed in recent years. The common feature of these quantum fluids is the weak interaction between the particles, which generally can be well described using mean field theories where the interaction is considered as a local, contact-like scattering \cite{Bloch08}. In contrast, cold dipolar fluids are composed of particles which carry a permanent electric dipole. Due to the strength and longer range of the dipole-dipole interaction, dipolar fluids are predicted to display physics that goes beyond a mean field description\cite{Lahaye09}.  In particular, cold dipolar bosons are expected to have new quantum as well as classical multi-particle correlation regimes\cite{Pupillo10,Laikhtman09,Lahaye09}. Observing the many-body correlations will open a window to the complex underlying physics that may drive the fluid into different theoretically proposed collective phases such as dipolar superfluids, dipolar crystals and dipolar liquids\cite{Astrakharchik07,Buchler07,boening11,Berman12}. In this paper we show experimental evidence of a few  correlation regimes of a cold dipolar exciton fluid, created optically in a semiconductor bilayer heterostructure. In the higher temperature regime, the average interaction energy between the particles shows a surprising temperature dependence which is an evidence for correlations beyond the mean field model. At a lower temperature, there is a sharp increase in the interaction energy of optically active excitons, accompanied by a strong reduction in their apparent population. This could be an evidence for a sharp macroscopic transition to a dark state as was suggested theoretically\cite{Combescot07}.

\begin{figure}
\includegraphics[width=0.5\textwidth]{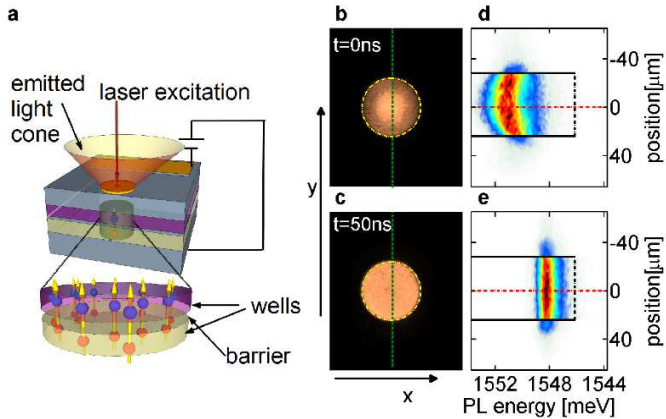}
\caption{\textbf{Dipolar excitons in an electrostatic trap: sample structure, experimental setup, and PL dynamics. a,} An illustration of the bilayer system, the dipolar excitons, and the circular electrostatic trap gate geometry. The excitation laser pulse impinges at the center of the trap. \textbf{b,c,d,e,} PL of an $X_{id}$ fluid inside an electrostatic trap, at two different times after a non-resonant excitation pulse. The first stage of the dynamics starts with a fast expansion of the dense and hot carriers due to the carrier-carrier repulsion (not seen here), followed by a cooling and a formation of $X_{id}$. Due to their strong dipole-dipole repulsion, these $X_{id}$ continue to expand rapidly towards the edges of the circular trap \cite{Chen06}, where they are confined through the interaction of their dipole with the externally applied electric field under the trapping gate.  \textbf{b},\textbf{c} Real space images of the $X_{id}$ fluid PL from an electrostatic trap during the laser pulse and 50ns after the laser pulse, respectively. The PL is spectrally filtered to collect only the emission from $X_{id}$ fluid. Note that the $X_{id}$ fluid reaches a homogeneous distribution in the trap in \textbf{c}. The dashed yellow line mark the trap gate boundary. \textbf{d},\textbf{e}, Spectral colormap images (in log scale) of the $X_{id}$ PL, taken from the cross-sections of the electrostatic trap shown by the green dash lines in \textbf{b},\textbf{c}, respectively. The dot-dashed red line marks the spatial location of the excitation spot, and the horizontal black lines mark the trap gate boundary. The vertical black dot-dashed line marks the energy at the bottom of the trap. The PL is clearly blue shifted with respect to this energy, due to mutual dipolar interactions between particles.}
\label{fig:basic}
\end{figure}

There are currently only a few feasible realizations of quantum dipolar fluids that are being experimentally tested. Perhaps the most known are dipolar atoms\cite{Lahaye09} or polar molecules \cite{Carr09} in either magneto-optical traps or optical lattices\cite{Bloch08}, and indirect dipolar exciton condensates in semiconductor quantum structures\cite{eisenstein04,High12}. Indirect dipolar excitons ($X_{id}$) are coulomb-bound electron-hole pairs inside an electrically gated semiconductor bilayer (also known as a double quantum well-DQW). $X_{id}$ are two-dimensional (2D) boson-like quasi-particles (see illustration in Fig.\ref{fig:basic}a) with 4 quasi degenerate spin states. The two states with spin $S=\pm$1 are optically active ("bright"), and the two states with spin $S=\pm$2 are optically inactive ("dark")\cite{Combescot07}. The $X_{id}$ carry a static electric dipole due to the separation of the electron and the hole into the two adjacent layers.  Furthermore, all the dipoles are aligned perpendicular to the layers, so that the dominant interaction between the $X_{id}$ is an extended repulsive dipole-dipole interaction\cite{Lee09, Schindler08}. The unique advantage of $X_{id}$ systems is that the effect of the interactions between the excitons can be observed directly: the interaction of  a given exciton with its surrounding excitons is manifested in an excess energy (called the "blue shift" - $\Delta E$), carried away from the system by a photon as the exciton recombines radiatively. It was suggested theoretically that this observed interaction energy could be used as a direct experimental probe of the various particle correlation regimes and the thermodynamic phases of $X_{id}$ systems\cite{Schindler08,Laikhtman09}, if it can be mapped as a function of the fluid temperature and density\cite{Stern08B}. However, calibrating the fluid density reliably at different temperatures turned out to be a non-trivial task in optically excited exciton systems\cite{Cohen11} which so far hindered direct and consistent observations of interaction-induced particle correlations.

Here we present time-resolved photoluminescence (PL) experiments of an optically excited $X_{id}$ fluid trapped inside an electrostatic trap\cite{Rapaport05,Hammack06,Schinner11}. We extract a consistent mapping of $\Delta E$ for a range of bright exciton densities ($n_b$) and temperatures.  We observe multi-particle correlations in the dipolar exciton fluid, and evidence for a macroscopic transition where the fluid redistribute its density with dark states which are uncoupled to light. Fig.~\ref{fig:basic}b and c show typical time resolved PL images of an $X_{id}$ fluid inside an electrostatic trap after its excitation with a non-resonant pulsed laser. About $50ns$ after excitation, the fluid reaches a dynamical equilibrium between the dipole-dipole repulsion of excitons that tends to drive the fluid outwards, and the confining "flat well" potential induced by the electrostatic gate. This equilibrium results in a uniform and homogeneous PL distribution inside the trap, indicating a flat density profile. This is clearly seen in Fig.~\ref{fig:basic}c. Fig.~\ref{fig:basic}d and e present the corresponding spatial-spectral images taken along the central axis of the trap gate. Fig.~\ref{fig:basic}e shows that the homogeneously distributed PL is blue shifted from the emission energy of a single exciton. This positive blue shift energy $\Delta E$ is due to the repulsive dipole-dipole interaction inside the $X_{id}$ fluid. In general, $\Delta E$ increases as $n_b$ increases and its value is sensitive to the intricate multi-particle correlation\cite{Schindler08,Laikhtman09}.

\begin{figure}
\includegraphics[width=0.5\textwidth]{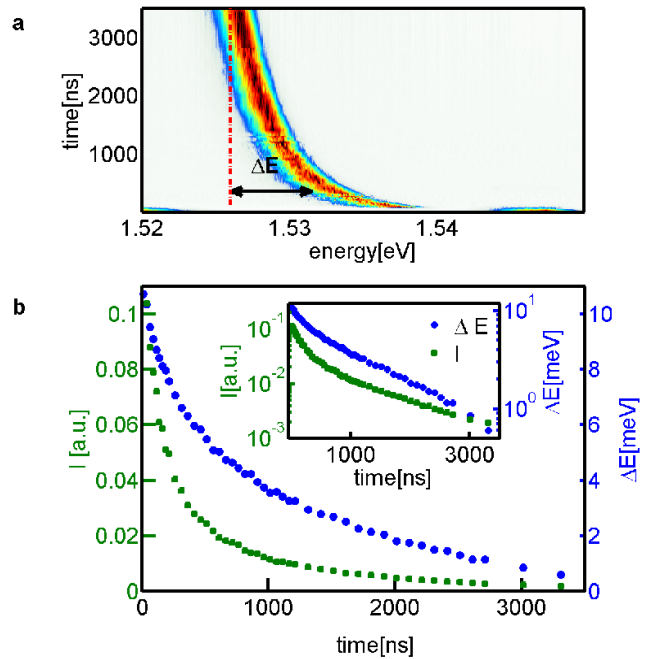}
\caption{\textbf{PL dynamics of a trapped $X_{id}$ fluid. a,} Spatially integrated PL  spectra of an $X_{id}$ fluid in a trap (taken at $3K$) at different times after the short excitation pulse. The spectra intensities are all normalized to unity for convenience.  The dot-dashed red line indicates the extrapolated $X_{id}$ energy as the density of the bright excitons goes to zero. The blue shift energy, $\Delta E$, is measured from this extrapolated energy, as is marked by the black arrow. \textbf{b,} The extracted time dependence $\Delta E$ (blue circles), and the integrated intensity $I$(green squares) from the $X_{id}$ spectra in \textbf{a}. It is seen that $\Delta E$ and $I$ decay non-exponentially and at different rates, due to the dependence of the effective $X_{id}$ lifetime on $\Delta E$ (see text). The inset presents the same data as in \textbf{b} but in log scale.}
\label{fig:data}
\end{figure}

\begin{figure*}
\includegraphics[width=1\textwidth]{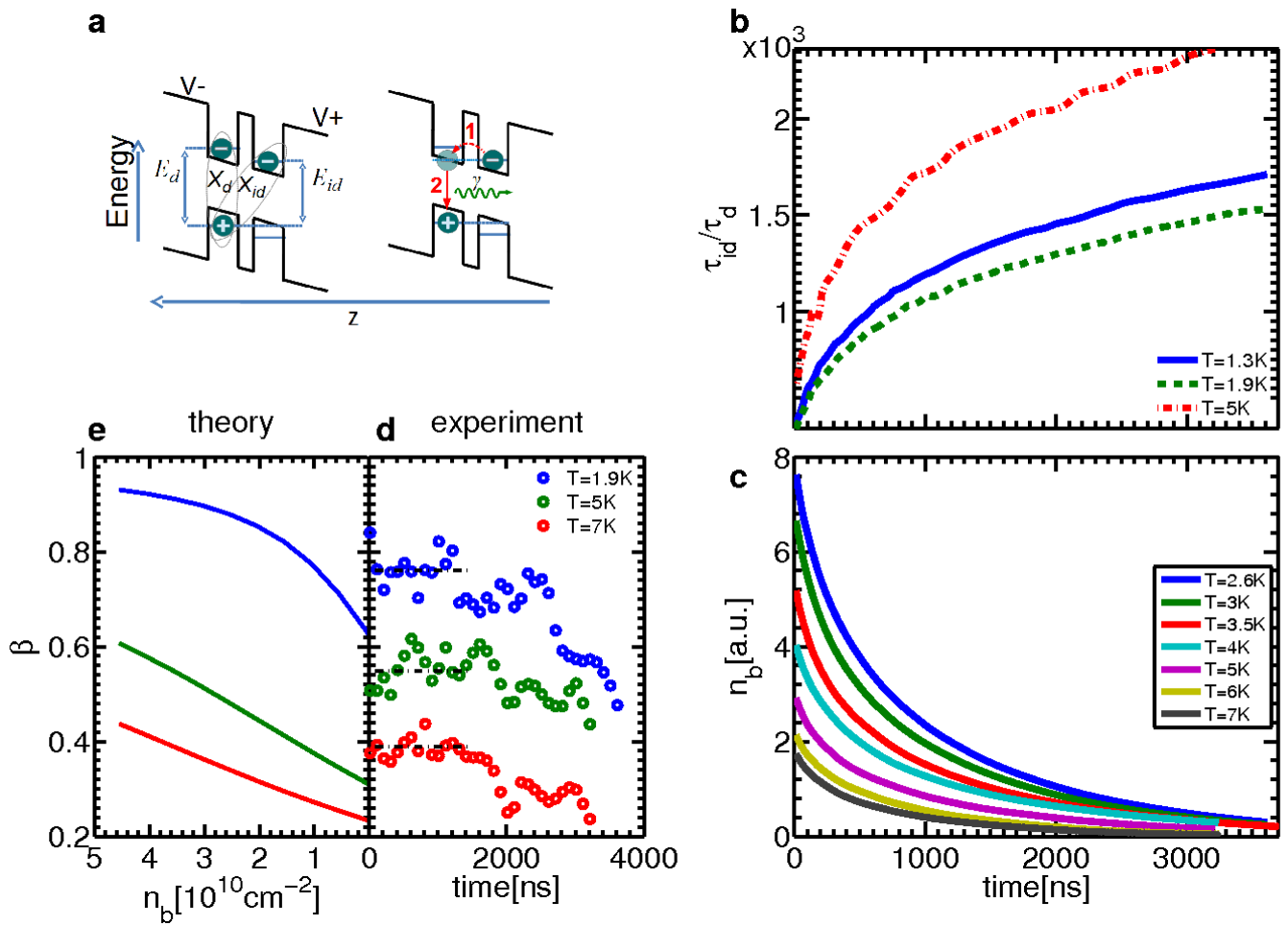}
\caption{\textbf{Optical recombination model, data analysis, and $X_{id}$  density calibration. a,} On the left side a schematic illustration of the energy band diagram of a DQW (in the growth direction) under an applied bias is shown . The energies of the direct exciton ($E_d$) and the dipolar exciton ($E_{id})$ are marked. The right side illustrates the process of an $X_{id}$ optical recombination, in which the electron effectively tunnels to the adjacent well (stage 1) and recombines with the hole (stage 2), emitting a photon. \textbf{b,} Extracted $\tau_{id}/\tau_{d}$ vs. time for three experimental temperatures $T=1.3K$, $T=1.9K$ and $5K$, using Eq.~(\ref{eq:tau}). \textbf{c,} The bright exciton density, $n_b(t)$, as a function of time for different temperatures, extracted using the calibration procedure described in the text. \textbf{d,} The experimentally obtained values of $\beta$ at different times for three different temperatures. \textbf{e,} Calculated values of $\beta$ as a function of $n_b$ for the same temperatures as in \textbf{d}, using an ideal 2D Bose-Einstein thermal distribution.
}
\label{fig:tau}
\end{figure*}

Fig.~\ref{fig:data}a presents an example of the spatially integrated and normalized PL spectra, taken at $T=3K$, at different times after the excitation pulse. The spectral position of the PL line shifts with time to lower energies as $n_b$ decreases. At long times, the PL energy asymptotically reaches a constant value. The difference between the PL energy at any given time to this asymptotic value, is the blue shift energy $\Delta E$ (marked by the arrow in Fig.~\ref{fig:data}a). The time dependence of the spectrally integrated PL intensity ($I$) and of $\Delta E$ are plotted in Fig.~\ref{fig:data}b. As the $X_{id}$ density drops with time, both $I$ and $\Delta E$ decreases with a non-exponential decay rate. The reason for this non-exponential decay is the dependence of the $X_{id}$ radiative recombination time ($\tau_{id}$) on $n_b$: as is illustrated in Fig.~\ref{fig:tau}a, radiative recombination of the $X_{id}$ can be described by a tunneling of either the electron or the hole (with a much lower probability due to its larger mass) to the adjacent well, where direct optical recombination with the opposite-charge particle takes place with a direct exciton recombination time - $\tau_d$. The tunneling probability depends on the difference between the direct and indirect transition energies, $E_d-E_{id}$. The larger this energy difference is, the larger is $\tau_{id}$ compared to $\tau_d$.
This picture can be quantified to get an expression for $\tau_{id}$ in terms of $\tau_d$ and $E_d-E_{id}$ [see supplementary material (S1) for more details]:
\begin{equation}\label{eq:tau}
    \frac{1}{\tau_{id}}=\frac{|c|^2}{\tau_d}=\frac{1}{\tau_{d}}\frac{v^2}{(E_d-E_{id})^2},
\end{equation}
where $|c|^2$ is the probability for an electron to tunnel to the hole QW, and $v$ is the tunneling matrix element. Note that while the non-polar, direct transition energy $E_d$ is independent of density, the dipolar energy $E_{id}$ depends on $n_b$. The time dependence of $\tau_{id}/\tau_d$ can be extracted from Eq.~\ref{eq:tau} by plugging in it the experimental values of $E_d-E_{id}(t)$. Fig.~\ref{fig:tau}b presents this time dependence for the three exemplary temperatures of $T=1.3K$, $1.9K$ and $5K$. Because the dominant $X_{id}$ recombination channel is radiative\cite{Rapaport04,Butov04prl,Sivalertporn12}, the dynamics of $n_b$, and its relation to the observed PL intensity $I$, can be described by a simple rate equation. Assuming an equilibrium of bright and dark $X_{id}$~\cite{Maialle93} (i.e., $n_b=n_d$ where $n_d$ is the dark $X_{id}$ density), we get:
\begin{multline} \label{eq:den}
    I(t)=-\alpha(T)\frac{d}{dt}(n_b(t)+n_d(t))=-2\alpha(T)\frac{dn_b(t)}{dt}=  \\
    =2\alpha(T)\frac{n_{rad}(t)}{\tau_{id}}=2\alpha(T)\frac{\beta(T)n_b(t)}{\tau_{id}},
\end{multline}
where $n_{rad}$ is the density of optically active excitons with in-plane k-vectors which are inside the radiation light cone,  $\beta\equiv (n_{rad}/n_b)$, and $\alpha(T)$ is the fraction of the total emitted photon flux that is collected by the detector (see \cite{Piermarocchi97}, and supplementary material for more information).
We now note that counting all the emitted photons from a given time $t$ after the excitation to $t\rightarrow\infty$ (where $n_b=0$) yields $n_b(t)$, i.e.,
\begin{equation}\label{eq:den2}
n_b(t)=\frac{\int_t^{\infty}I(t')dt'}{2\alpha(T)}
\end{equation}
This means that their densities are equal so that $n_b$ is only half of the total exciton density.
Secondly, combining Eq.~(\ref{eq:den}) with Eq.~(\ref{eq:den2}) we get a relation between $I(t)$, $\tau_{id}(t)$, and $\beta(T)$:
\begin{equation}\label{eq:int}
    I(t)=\frac{\beta(T,t)\int_t^{\infty} I(t')dt'}{2\tau_{id}(t)}.
\end{equation}
Since $\tau_{id}$ was extracted \textit{independently} from the PL \textit{energy} using Eq.~\ref{eq:tau}, comparing the two sides of the equation yields $\beta(T,t)$. This dependence is plotted for 3 different temperatures in Fig.~\ref{fig:tau}d. $\beta$ increases with decreasing time, i.e., with increasing $n_b$. Also, $\beta$ decreases with temperature. This density dependence is a signature of a deviation from a pure classical ideal gas distribution.  Fig.~\ref{fig:tau}e plots the theoretically calculated values of $\beta(n_b)$ for the 3 corresponding temperatures, using an ideal 2D Bose-Einstein (BE) model [see supplementary material (S3) for the full derivation]. There is a reasonable qualitative agreement between the calculation and the experiment, indicating the validity of the model assumptions. Note however that currently we cannot obtain a direct comparison between the theoretical and experimental values of $\beta$ as no absolute calibration of $n_b$ exists. Another strong verification for the validity of the above analysis was done for a trapped $X_{id}$ fluid in steady state under CW excitation and is shown in the supplementary material. Fig.~\ref{fig:beta}a shows in green circles the temperature dependence of $\beta$ at the high density limit (marked by the black dashed lines in Fig.~\ref{fig:tau}d. $\beta(T)$ increases as $T$ decreases down to $\sim$2.5K, where it suddenly drops. This behavior is fitted to an ideal BE distribution, shown by the solid blue line. For temperatures above $\sim$2.5K the theoretical prediction fits well with the experimental data. This means that for $T\gtrsim$2.5K, the $X_{id}$ fluid has a well defined thermal distribution, but sharply deviates from it at lower temperatures. This is the first important observation of this analysis.

Next we would like to map the dependence of $\Delta E$ on $T$ and $n_b$.  This can be done with a common experimental calibration for the optically-active exciton densities for all temperatures using Eq.~(\ref{eq:den2}). To do this in a simple tractable manner, we calculate an approximate, density independent value of $\alpha(T)$. We can then use this calculated value with Eq.~(\ref{eq:den2}) and the experimental values of $I(t)$ to get $n_b(t)$ for each $T$. The results are plotted in Fig.~\ref{fig:tau}c. This procedure allows us to compare the behavior of the $X_{id}$ fluid with similar densities but at different temperatures.

\begin{figure}
\includegraphics[width=0.5\textwidth]{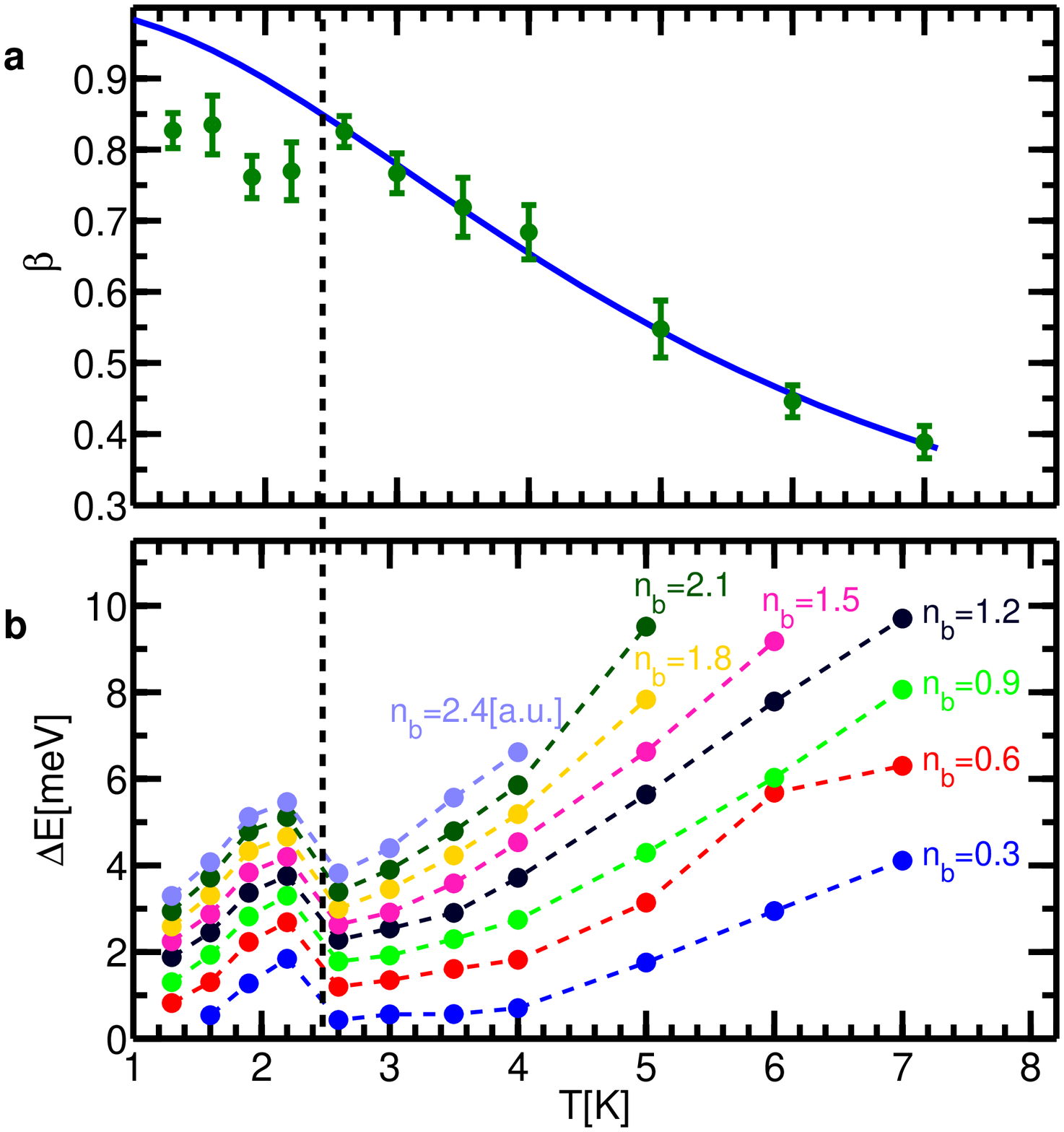}
\caption{\textbf{ Particle correlation regimes. a,} $\beta$ values at the high exciton density limit (marked by the black dashed lines in Fig.~\ref{fig:tau}d), as a function of $T$ (green circles). The solid blue line is the theoretical values of $\beta$ assuming an ideal 2D Bose-Einstein thermal distribution with $n_b=3.5\cdot 10^{10}[cm]^{-2}$. \textbf{b,} $\Delta E$ as a function of $T$ for different values of bright exciton densities, $n_b$ (dashed lines are guides to the eye). The vertical black dashed line mark $T_c$, the boundary between the two regimes as is discussed in the text. A lower bound for $n_b$ can be obtained from the blue shift at the highest temperature by applying the mean field model \cite{Laikhtman09} yielding $n_b\geq 2.2\cdot 10^{10}cm^{-2}/1(a.u.)$. For this density estimate we assume that at the highest temperature, the bright and dark exciton densities are identical and therefore $n_b$ is half of the total particle density.
}
\label{fig:beta}
\end{figure}

\begin{figure}
\includegraphics[width=0.5\textwidth]{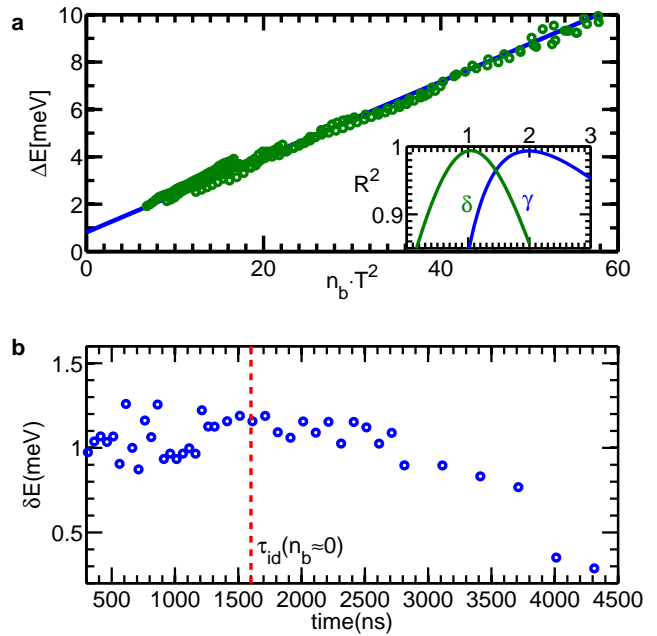}
\caption{\textbf{ Data analysis. a,} Scaling of the data of $\Delta E$ in \ref{fig:beta} to $n_bT^2$ for all $T>T_c$. The solid straight line was added as a guide to the eye. The inset shows the $R^2$ values of the quality of scaling of the experimental $\Delta E$ data to $n_b^\delta T^\gamma$ for different values of the exponents $\gamma$ and $\delta$. \textbf{b,} The time dependence of the magnitude of energy "jump" given by $\delta E=E_{id}(T=2.2K)-E_{id}(T=2.6K)$, where these two temperatures correspond to the temperatures just below and above $T_c$ respectively. The dashed red line marks the value of $\tau_{id} (T=2.2K)$ at the low density limit.
}
\label{fig:scaling}
\end{figure}

Fig.~\ref{fig:beta}b presents the experimental dependence of $\Delta E$ on $T$ for different fixed densities. Two distinct temperature regimes are observed for all densities, corresponding exactly to the two regimes seen for $\beta(T)$, with a sharp transition between them at $T_c\simeq$2.5K. For all temperatures above $T_c$, a clear temperature dependence of $\Delta E$ is observed. $\Delta E$ decreases with decreasing $T$. This dependence is a clear evidence for particle correlations beyond mean field. In contrast, a mean field calculation of $\Delta E$ predicts a "capacitor formula" dependence that is temperature independent\cite{Butov99}.  As the dipole-dipole interaction between the excitons is repulsive, a reduction of $\Delta E$ for a given density $n_b$ means an increase in the particle correlations: the more the $X_{id}$ spatially correlate to minimize their energy, the smaller $\Delta E$ will be. Therefore, the results suggest that as $T$ decreases, the spatial correlations of the excitons in the fluid increase. To better quantify the dependence of $\Delta E$ and therefore the particle correlations on $n_b$ and $T$ in this regime, we look for a scaling law of our data. Fig.~\ref{fig:scaling}a plots $\Delta E$ for a large set of densities and for all the measured temperatures above $T_c$, as a function of $n_b T^2$. The data collapse into a single linear line to a high accuracy (see inset). The linear dependence of $\Delta E$ on $n_b$ suggests a lack of long range order in the fluid\cite{Laikhtman09}. The scaling of $\Delta E$ on $T^2$ is surprising. In contrast, the models of Refs.\cite{Schindler08,Laikhtman09} predict a much weaker, sublinear dependence of $\Delta E$ on $T$, if the dipoles are a classically correlated gas. This specific temperature dependence could be an indication for a transition of the fluid correlations from classical to quantum. While the former are expected to lead to a clear temperature dependence of $\Delta E$, the latter should have a much weaker dependence, as was calculated in Ref.~\cite{Laikhtman09}. This transition to a temperature independent $\Delta E$ is especially clear for the low densities of Fig.~\ref{fig:beta}b, and it happens at a temperature range very similar to the one where quantum degeneracy of $X_{id}$ was reported very recently \cite{High12}. Lower bound estimation for the $X_{id}$ density (see caption of Fig~\ref{fig:beta}b) indeed suggests that the $X_{id}$ fluid should become quantum degenerate (see Fig S3 in the supplementary material) for all the densities presented.

Turning to the other regime, we observe a sharp increase in $\Delta E$ for all densities just below $T_c$. This jump correlates well with the onset of the
deviation from the theoretical values of $\beta$ plotted in Fig.~\ref{fig:beta}a, where we observe a sharp drop of $\beta(T<T_c)$ with much \emph{less} radiative $X_{id}$ than the theoretical prediction of a BE gas of bright excitons (plotted in blue). In other words, suddenly below $T_c$ there seem to be less bright excitons but yet more interaction energy. This is an indication for a sudden and sharp depletion of the bright exciton density and a sudden macroscopic transition to an optically inactive "dark" state below $T_c$. This increase in the density of the dark state is seen in $\Delta E$ of bright excitons as these dark excitons still interact with the bright excitons. A Bose-Einstein condensation (BEC) of dark excitons and its effect on the excited bright exciton energy was recently suggested in a theoretical paper by Combescot \emph{et. al.}\cite{Combescot07}. It was proposed that in perfect excitonic systems, the dark excitons should have an energy slightly lower than the bright excitons, and therefore at low enough temperatures and high densities, a BEC should form in the dark state. In practice however, due to disorder and dipolar interactions, it is more likely that the ground state consists of a mixture of the bright and dark excitonic states. The following possible scenario is therefore consistent with our experimental observations: for all temperatures, the pulse excitation creates a large density of hot particles that very quickly (within a few nanoseconds) cool down to the lattice temperature. For $T>T_c$, due to efficient spin flip processes between dark and bright states\cite{Maialle93,Leonard10}, their population is approximately equal throughout the optical recombination process and their density decay together with time. At temperatures below $T_c$, the high density fluid cools down and condenses fast after excitation, pulling bright excitons to the dark ground state so that the population equality between the two species breaks down, resulting in more dark excitons and less bright excitons than expected, as is seen in Fig.\ref{fig:beta}a,b. The fact that the temperature dependence of this transition is very sharp (a fraction of a Kelvin), excludes the possibility of a simple thermal re-population of a lower dark state, but rather indicates to a sharp macroscopic transition. After the condensation, it is expected that the scattering between the condensed particles in the fluid will be strongly suppressed, leading to a suppression of spin flip processes and therefore to an effective decoupling of the dark $X_{id}$ from the bright ones. Such scattering suppression was recently observed and analyzed theoretically \cite{High12}. Since the condensation and the bright-dark decoupling happens shortly after the pulsed excitation, it should be hard to directly observe the existence of a dark state by monitoring the dynamics of the bright $X_{id}$ PL intensity alone. However,  there is a way to probe the dark state existence, as can be seen from Fig~\ref{fig:scaling}b. Here we plot the time dependence of the energy "jump" given by $\delta E=E_{id}(T=2.2K)-E_{id}(T=2.6K)$, where these two temperatures correspond to the temperatures just below and above $T_c$ respectively. It can be seen that $\delta E$ persists for times much longer than the bright exciton lifetime (marked by the red dashed line), which indicates that there is a dark state in the system affecting the energy of the bright $X_{id}$ via mutual dipolar interactions. As can be seen in this figure, this state is populated for times much longer than the longest lifetime of the bright excitons, as is expected from a dark excitonic state which is weakly coupled to light.

To summarize, the above results show a few distinct correlation regimes of a 2D dipolar exciton fluid. We note that due to the complexity of the system and the inherent problems of measuring a dark state directly, a consistent theoretical framework that can describe these effects as well as further experimental efforts are therefore an outstanding challenge.

We would like to thank Oded Agam, Paulo Santos, and Snezana Lazic for useful discussions. This work was partially supported by DFG Project No.
581021. And by the Israeli Science Foundation Project No. 1319/12.

\end{bibunit}

\onecolumngrid
\newpage
\newcommand{\sgn}{{\rm sgn}}

\renewcommand*{\theequation}{S\arabic{equation}}
\renewcommand*{\thefigure}{S\arabic{figure}}
\renewcommand*{\thesubsection}{S\arabic{section}}
\renewcommand*{\citenum}{S\arabic{cite}}

\renewcommand{\bibnumfmt}[1]{#1.}

\setcounter {equation} {0}
\setcounter {figure} {0}
\setcounter {page} {1}

\begin{center}
{\Large  \textbf{Supplementary material}}
\end{center}

\begin{normalsize}
\begin{spacing}{1.5}
\begin{bibunit}
\section{Experimental details}

The sample that is used in the experiment is an MBE grown bilayer structure consisting of a $120/40/120{\AA}$ - $GaAs/Al_{0.45}Ga_{0.55}As/GaAs$ DQW on top of a n-doped GaAs substrate which serves as a bottom electrode. A semi-transparent metallic ($Ti$) circular electric gate, with a 50$\mu m$ diameter, is micro-fabricated on top of the structure, and is connected to a top electrode, as illustrated in Fig~1a. The area of the circular gate forms an electrostatic trap for the $X_{id}$ \cite{Hagn95}, which remain confined under it. The DQW structure is placed much closer to the bottom electrode than to the top gates, to prevent a significant charge separation that can occur on the boundary of the trap \cite{Rapaport05,Hammack06,Kowalik-Seidl12}. The sample is mounted into a liquid $^4$He optical cryostat. The sample temperature in these experiments was varied in the range of 1.3-7K. The sample is excited non-resonantly with a $671$nm Q-switched laser with a pulse duration of 15ns and a repetition rate of $25$kHz, focused on the center of the trap gate. The time and spatially resolved spectral images following the excitation pulses are collected by a fast gated intensified CCD camera (PIMAX-II) mounted on a spectrometer.

\section{Exciton life time in double well system}

The recombination time of direct excitons is a few hundreds of picoseconds. The recombination time of dipolar excitons can be a few orders of magnitude larger. The reason for this difference is that for exciton recombination to happen, the electron and hole forming an exciton have to be located at the same place while electrons and holes forming dipolar excitons are located in different quantum wells. For recombination an electron has to tunnel from the electron well to the hole well. Hole tunneling to the electron well is negligible compared to the electron tunneling.

If the tunneling is neglected then an electron - hole pair can be in one of two
possible states: a direct exciton and an indirect exciton, excited states are not
important for recombination. The electron functions in the two different wells are not
orthogonal and their overlap controls the tunneling probability. The
recombination of an indirect exciton can be considered as tunneling to a
virtual direct exciton state and recombination of this direct exciton. The
difference between the energies of direct and indirect exciton is negligible
compared to the energy of a photon emitted in the recombination event.
Therefore the recombination time of indirect excitons, $\tau_{id}$, differs
from the recombination time of direct excitons, $\tau_{d}$, by a small
probability $w_{t}$ of the tunneling between the two exciton states:
\begin{equation}
\frac{1}{\tau_{id}} = \frac{w_{t}}{\tau_{d}} \ .
\label{eq:1}
\end{equation}

For calculation of $w_{t}$ we use the model of two wells of the same width
separated by a barrier and assume that the ground state of an electron in one
of the wells is perturbed by: (a) the overlap with the ground wave function in the
other well, (b) the electric field applied perpendicular to the wells, (c) the
attraction to the hole to which it is bound, and (d) the interaction with electrons and
holes belonging to different excitons. The main assumption is that all these
perturbations are small compared to the quantization energy of an electron in a
single well, or to the energy separation between the ground state and the first
excited state. The electron wave function is approximated by a linear combination
of the ground wave functions in the separate wells,
\begin{equation}
\psi(x) = c_{l}\psi_{l}(x) + c_{r}\psi_{r}(x) \ .
\label{eq:2}
\end{equation}
Neglecting an admixture of excited states we neglect a change of the
exponential decay of the wave functions under the barriers due to perturbation
of the ground state energies (To take into account this change it is possible
also to include the energy correction in the original approximation). A system
of equations for the coefficients $c_{l}$ and $c_{r}$ is
\begin{subequations}
\begin{eqnarray}
&& (E_{0} + v_{ll} + eFx_{ll} + V_{int,ll}) c_{l} +
    (E_{0}u + v_{lr} + eFx_{lr} + V_{int,lr}) c_{r} =
    E(c_{l} + uc_{r}) \ ,
 \\
&& (E_{0}u + v_{rl} + eFx_{rl} + V_{int,rl}) c_{l} +
    (E_{0} + v_{rr} + eFx_{rr} + V_{int,rr}) c_{r} =
    E(uc_{l} + c_{r}) \ .
%\label{eq:3}
\end{eqnarray}
\label{eq:3}
\end{subequations}
where $E_{0}$ is the energy of the unperturbed ground state, $v_{ij}$ is the
matrix element of the perturbation of the two-well structure compared to one
well, $F$ is
the electric field, $x_{ij}$ the coordinate matrix element, $V_{int,ij}$ is the
matrix element of the interaction energy, and $u$ is the overlap integral of
$\psi_{l}$ and $\psi_{r}$. Continuing in the frame of perturbation theory we
retain only the off-diagonal matrix elements of the structure perturbation
$v\equiv v_{rl}=v_{lr}$ that do not have another small parameter. Transmission
of an electron from one well to another changes only its energy in the electric
field and the interaction energy. Therefore Eq.(\ref{eq:3}) is reduced to
\begin{subequations}
\begin{eqnarray}
&& \left(E_{1} - E - \frac{eFd}{2} + \frac{V_{int,d}}{2}\right) c_{l} +
    v c_{r} = 0 \ ,
\label{eq:4a} \\
&& v c_{l} + \left(E_{1} - E + \frac{eFd}{2} - \frac{V_{int,d}}{2}\right)
    c_{r} = 0 \ ,
\label{eq:4b}
\end{eqnarray}
\label{eq:4}
\end{subequations}
where $E_{1}$ includes $E_{0}$ and all corrections that are the same in both
wells, $d$ is the separation between the centers of the wells and $V_{int,d}$
is the difference of the electron interaction energy in the two wells. This energy
includes the binding energy difference between direct $E_{d}$ and indirect
$\widetilde{E_{id}}$ excitons and also the (doubled) interaction energy with other excitons (the
interaction energy changes its sign as a result of electron transmission
between the wells). The electrostatic energy $eFd$ is an addition to the binding
energy of indirect exciton. Solving Eqs.(\ref{eq:4}) leads to the following
expression for the amplitude of the electron wave function in the hole well
\begin{equation}
c = \frac{1}{\sqrt{2}}
        \left[
    1 - \frac{E_{d} - \widetilde{E_{id}} - eFd + \Delta E}
    {\sqrt{(E_{d} - \widetilde{E_{id}} - eFd + \Delta E)^{2} + 4v^{2}}}
        \right]^{1/2} ,
\label{eq:5}
\end{equation}
where $\Delta E$ is the interaction energy between indirect excitons. The
tunneling probability is
\begin{equation}
w_{t} = |c|^{2} \ .
\label{eq:6}
\end{equation}
If the tunneling matrix element $v$ (in our sample can be estimated by $v \approx$ 0.25meV) is small compared to the energy difference
(in our experiment its minimum value is 5meV)
then the expression for the tunneling probability is reduced to
\begin{equation}
w_{t} = \frac{v^{2}}{(E_{d} - E_{id})^{2}} \ .
\label{eq:7}
\end{equation}
were $E_{id}=\widetilde{E_{id}} + eFd - \Delta E$.
Substitution of Eq.(\ref{eq:7}) in Eq.(\ref{eq:1}) results in
\begin{equation}
\frac{1}{\tau_{id}} = \frac{1}{\tau_{d}} \
    \frac{v^{2}}{(E_{d} - E_{id})^{2}} \ .
\label{eq:8}
\end{equation}
 Which is identical to Eq.(1) in the main text.

\section{CW experiment}
In the main text the relation between $\tau_{id}$ and $\Delta E$ was derived and used (Eq.(1) and Eq.(\ref{eq:8})) for the data analysis. In order to further verify this analysis, the same model was used for the results of trapped $X_{id}$ fluid in steady state. In this experiment, the trapped $X_{id}$ was excited using a CW HeNe laser at T=5K. The corresponding $\Delta E$ for different laser powers ($G$) is presented in Fig.~\ref{fig:CW} by the blue circles. In steady state, were the number of the generated excitons equals to the number of excitons that recombine, we have that $n_b = G\times\tau_{id}$. Now we assume that $\Delta E \propto n_b$, a relation that was proven experimentally in the paper. By using Eq.~(1) from the main text, we can get the functional dependence of $G$ on $\Delta E$:
\begin{equation}\label{eq:CW}
    G = \frac{n_b}{\tau_{id}} \propto \frac{\Delta E}{(E_{id} - E_{d})^{2}}.
\end{equation}
\begin{figure}[h!]
\includegraphics[width=0.5\textwidth]{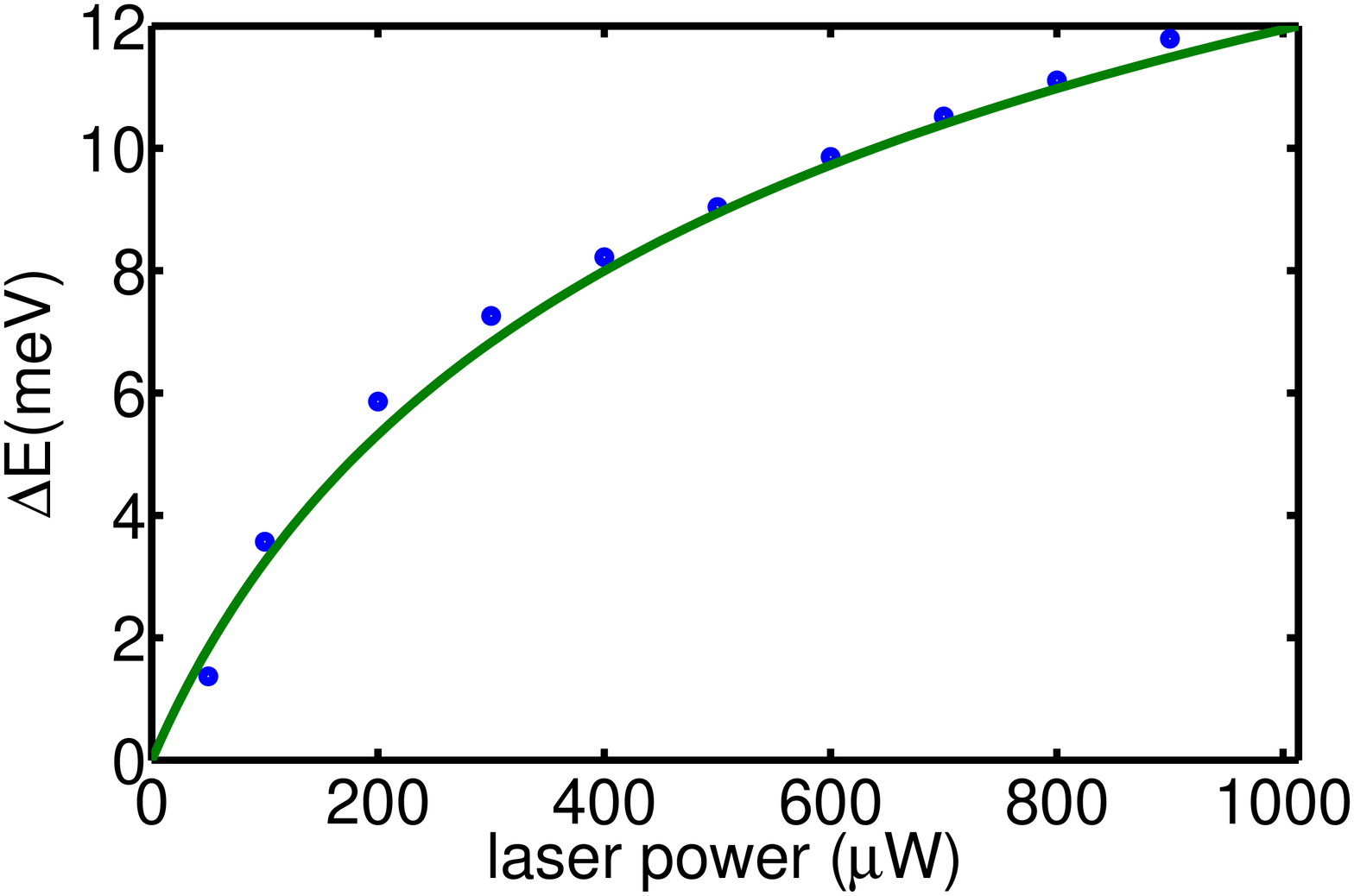}
\caption{(color online).
The dependence of $\Delta E$ on laser power (blue circles), in an exciton trap as the one discussed in the paper but under CW laser excitation instead of a pulsed one.
}
\label{fig:CW}
\end{figure}

The solid green line in Fig.~\ref{fig:CW} is a fit to Eq.~(\ref{eq:CW}) where only the proportionality constant is used as a fit parameter. A very good agreement between the model and the experiment is found, which adds another confirmation to our model.

\section{The influence of the light Cone on the radiation}
The energy of a dipolar exciton is given by $E_{ex}= E_0+E_{kinetic}$ where $E_0$ is the energy of a motionless exciton and the kinetic part is
$E_{kinetic}=\hbar^2 \vec{k}_\parallel^2/2m$ (the excitons can only move in the DQW plane). The photon energy inside the heterostructure is
$  E_{ph}=\hbar \frac{c}{n} |\vec{k}|$
where $\vec{k}=\vec{k}_\parallel+\vec{k}_\perp$ is the wave vector and $n$ is the effective refractive index. Since there is a translational symmetry in the DQW plane, $\vec{k}_\parallel$ is a conserved quantity. Requiring energy conservation of the exciton and the photon, we get the relation
\begin{equation}\label{eq:Energy_equation}
  E_0+\frac{\hbar^2 \vec{k}_\parallel^2}{2m}=\hbar \frac{c}{n} \sqrt{\vec{k}_\parallel^2+\vec{k}_\perp^2}.
\end{equation}
(\ref{eq:Energy_equation}) has of course numerous solutions, and the propagation angle of the emitted photon (see Fig. \ref{fig:angle}) is given by
 $ |\vec{k}_\parallel| / |\vec{k}|=\sin{\theta} .$
\begin{figure}[h!] \label{fig:emission}
  \centering
  \includegraphics[width=0.4\textwidth]{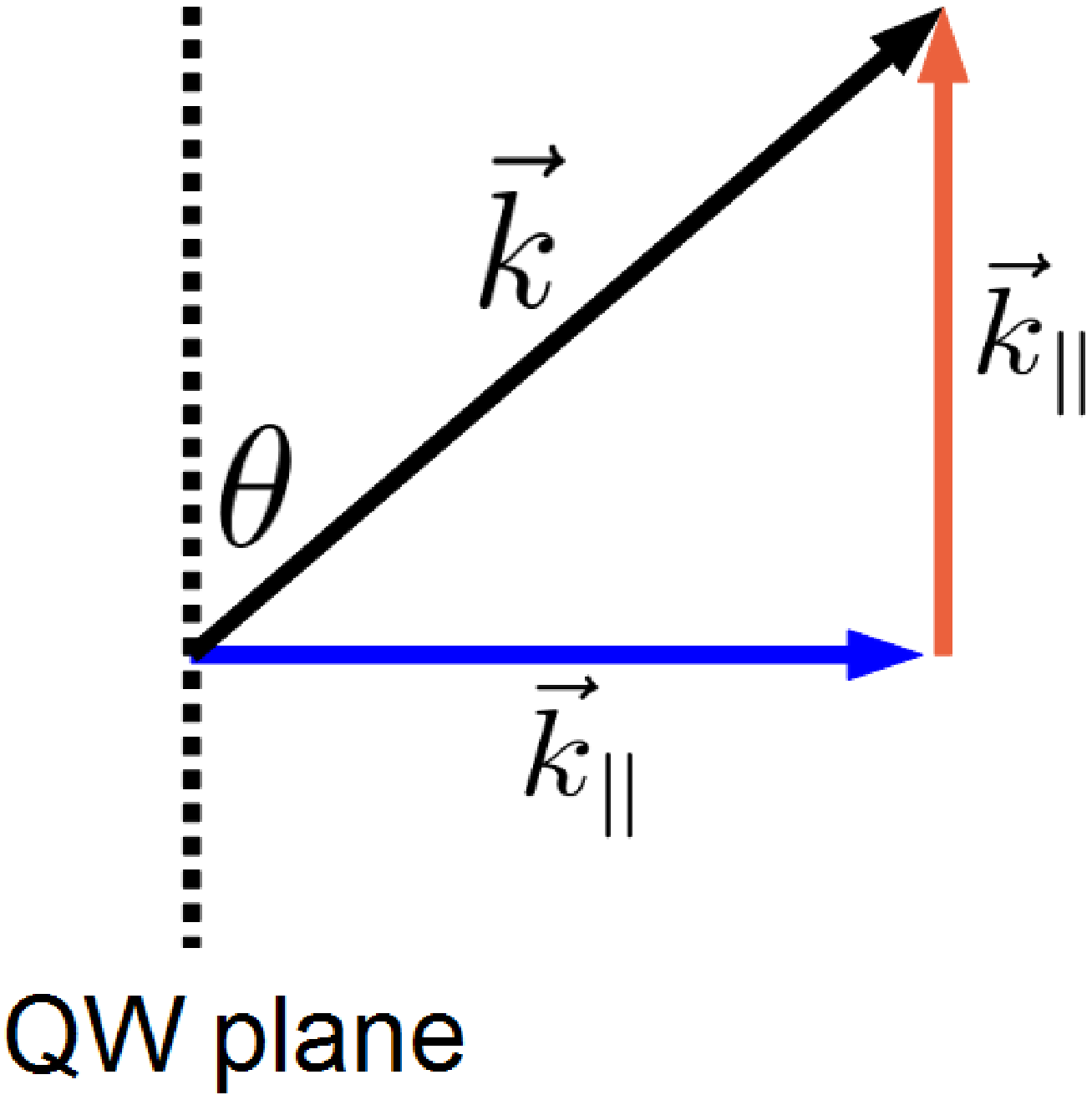}\\
  \caption{Parallel and perpendicular momentum of the emitted photon.}\label{fig:angle}
\end{figure}
There are two limiting cases for Eq. (\ref{eq:Energy_equation})
\begin{enumerate}
  \item {If $\vec{k}_\parallel=0$, then the photon is emitted perpendicular to the DQW plane.}
  \item {If $\vec{k}_\perp=0$, then the photon is emitted parallel to the DQW plane. This case yields the maximal $\vec{k}_\parallel$, which defines the \emph{light cone}.}
\end{enumerate}
Proceeding with the latter case and denoting $k_\parallel^*=|\vec{k}_\parallel|$, one gets
\begin{equation}\label{eq:k_light_cone}
  k_\parallel^*=\frac{m_{ex}}{\hbar}\left[\frac{c}{n}\pm\sqrt{\left(\frac{c}{n}\right)^2-\frac{2E_0}{m}}\right]
\end{equation}
Plugging in numbers: $E_0 \simeq 1515meV$, $m_{ex}\simeq 0.18m_0$ ($m_0$ is the free electron mass) and $n_{GaAs}\simeq 3.6$, one gets $k_\parallel^* \cong 2.8\cdot10^5 cm^{-1}$. Only excitons with $|\vec{k}_\parallel| \le k_\parallel^*$ can recombine and decay radiatively.

The energy difference $\delta E$ between excitons with $k_\parallel=0$ and $k_{\parallel}^*$ is
$\delta E=\hbar^2 {k_\parallel^*}^2/2m\approx 160 \mu eV$. Note that $1 ^\circ K\approx 90\mu eV$, so for thermal equilibrium at few $^\circ K$ there is a relatively large number of excitons outside the light cone. Only excitons with energy smaller than $\delta E$ can recombine, and their number is given by
$n_{rad}(T,n)=\int_0^{\delta E} f(\epsilon(T,n))g(\epsilon)\text{d}\epsilon$ where $f(\epsilon(T,n))$ is the energy distribution function and $g(\epsilon)$ is the density of states (which is a constant in 2D).

If we assume a thermal equilibrium and a classical Maxwell-Boltzmann (MB) distribution $f(\epsilon(T))\propto\exp(-\epsilon/kT)$, the fraction of optically active particles is given by
\begin{equation}\label{eq:n_frac1}
   \beta(T)\equiv \frac{n_{rad}(T)}{n_b}=\frac{\int_0^{\delta E} f(\epsilon)g(\epsilon)\text{d}\epsilon}{\int_0^\infty f(\epsilon)g(\epsilon)\text{d}\epsilon}=1-e^{-\delta E/kT}
\end{equation}

For an ideal Bose-Einstein distribution in 2D ,$\beta$ becomes density dependent. Assuming an equal number of bright and dark excitons, we plot the calculated temperature dependence of $\beta$ for different particle densities in Fig. \ref{fig:beta}. For low densities ($n_b\leq 10^{9}[cm]^{-2}$), $\beta$ coincides with the $\beta$ based on MB distribution as expected, and slightly increases for higher densities (by less then a factor of 2).

\begin{figure}[!ht]
  \centering
  \includegraphics[width=0.6\textwidth]{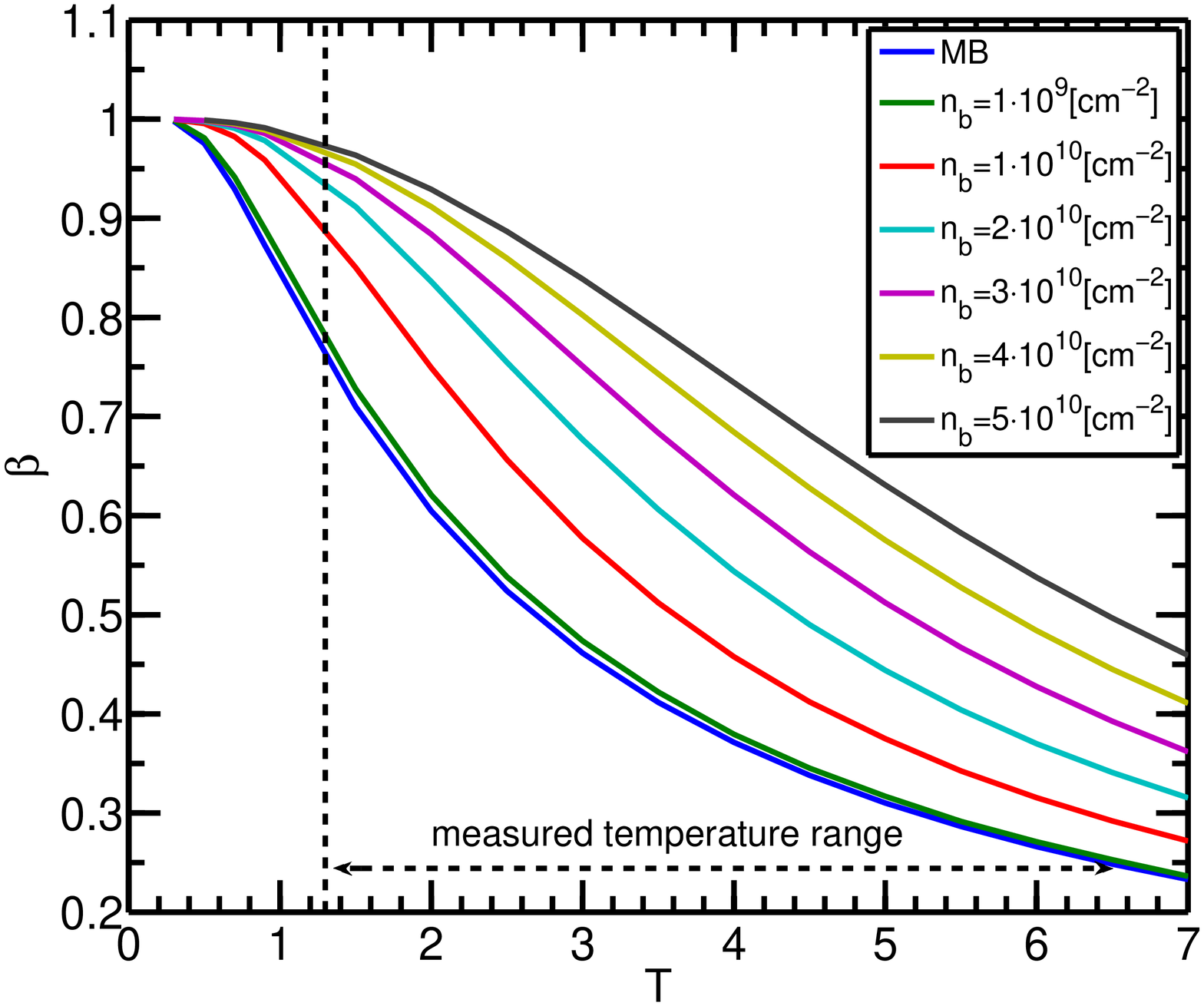}\\
  \caption{$\beta$ as a function of $T$ for MB statistics(blue line), and for BE statistics at different densities.}
  \label{fig:beta_s}
\end{figure}

In reality, the collected photons are only part of the radiative photons due to the numerical aperture (NA) of our experimental setup and the refraction of light going out from the sample (see Fig. \ref{fig:numerical_aperture}).
The PL intensity ($I$) that is collected by our apparatus is proportional to the decay rate of excitons: $ I=\alpha(T,n)\frac{dn_b}{dt}$. If we had a perfect lens which collects every emitted photon, this proportionality constant will be $\alpha=1$.
\begin{figure}[!ht]
    \centering
    {
        \includegraphics[width=0.5\textwidth]{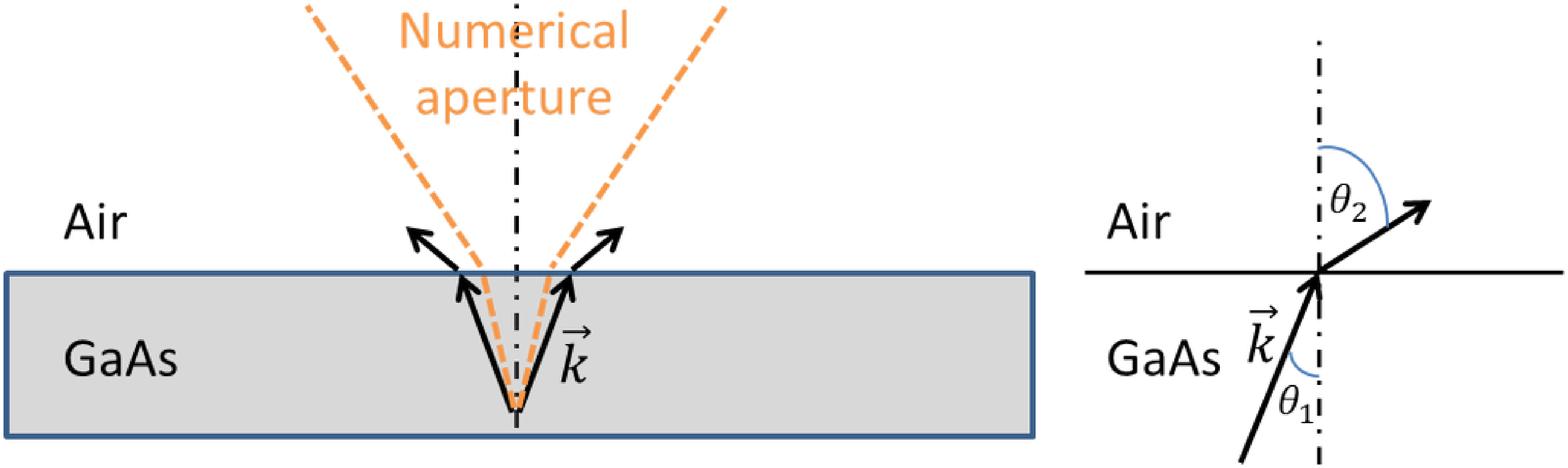}\\
        \caption{The influence of numerical aperture and the refraction of light on the collection efficiency}\label{fig:numerical_aperture}
    }
\end{figure}
If a photon with $k_\parallel$ is collected by our optical system, Snell's law requires that
\begin{equation}\label{eq:NA}
    n_{GaAs}\sin(\theta_1)=n_{GaAs}\frac{k_\parallel}{k}=\sin(\theta_2)\le NA
\end{equation}
For our experimental setup $NA=0.42$, and we can substitute $k=n_{GaAs}k_0$ in the above expression ($k_0$ is the wave vector magnitude in vacuum), so the maximal $k_\parallel$ that can be collected is $ k_\parallel^{col}=NA \cdot k_0\approx 0.42\cdot \frac{2\pi}{810nm}\approx 3.2\cdot10^4cm^{-1}$. This limit also implies an energetic limit, which we denote by $\delta E_{col}$. Note that $\frac{k_\parallel^{col}}{k_\parallel^*}\approx0.1$ and thus
\begin{equation}\label{eq:lightcone_E_fraction}
   \frac{\delta E_{col}}{\delta E}=\left(\frac{k_\parallel^{col}}{k_\parallel^*}\right)^2\approx0.01,
\end{equation}
which is the reason why the experimentally accessible part of the dispersion curve of excitons is essentially flat (of the order of $1\mu eV)$. Now, by assuming again thermal equilibrium, we can calculate the following:
\begin{equation}\label{eq:alpha(T)}
    \alpha(T,n_b)=\frac{\int_0^{\delta E_{col}} f(\epsilon,n_b)g(\epsilon)\text{d}\epsilon}{\int_0^{\delta E} f(\epsilon,n_b)g(\epsilon)\text{d}\epsilon}
    =\frac{\int_0^{\delta E_{col}} f(\epsilon,n_b)g(\epsilon)\text{d}\epsilon}{n_b\beta(T,n_b)}.
%    \xrightarrow{\Delta E\ll kT} \frac{\Delta E_{col}}{\Delta E}=const
\end{equation}
This complicated equation cannot be solved without a good knowledge of the real distribution function, which we do not have. However, it can be replaced with a simpler analytic expression if we assume MB distribution, yielding:
\begin{equation}\label{eq:alpha(T)MB}
    \alpha(T)=\frac{1-e^{-\delta E_{col}/kT}}{\beta(T)}=\frac{1-e^{-\delta E_{col}/kT}}{1-e^{-\delta E/kT}}
\end{equation}
This simplification gives a density independent lower bound for $\alpha(T,n_b)$. As the experimentally observed and calculated changes of $\beta$ with density are not  large, this simplifications should give a fairly good approximation to $\alpha$. This assumption might result in a possible underestimate of $\alpha$ mostly for very high densities (short times after the excitation), however, as can be seen in Fig. \ref{fig:alpha_T}, the values of $\alpha(T)$ do not depend strongly on the temperature in this range, and therefore small modification of these values should not have a significant effect on the results presented.
\begin{figure}[!ht]
  \centering
  \includegraphics[width=0.3\textwidth]{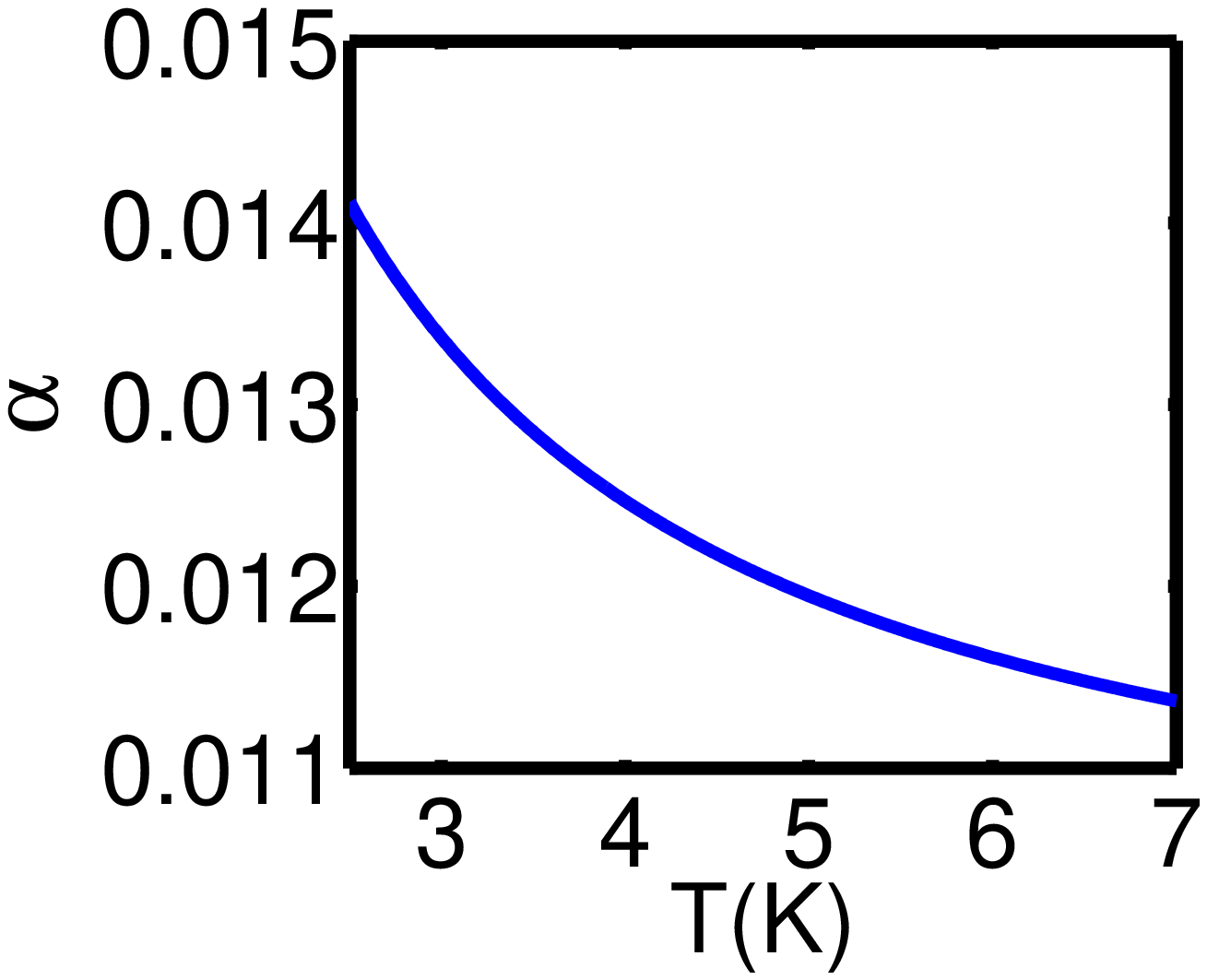}\\
  \caption{Temperature dependence of the collection efficiency $\alpha(T)$}\label{fig:alpha_T} calculated using Eq.~\ref{eq:alpha(T)MB}
\end{figure}

\end{bibunit}
\end{spacing}
\end{normalsize}


\begin{thebibliography}{28}%
\makeatletter
\providecommand \@ifxundefined [1]{%
 \@ifx{#1\undefined}
}%
\providecommand \@ifnum [1]{%
 \ifnum #1\expandafter \@firstoftwo
 \else \expandafter \@secondoftwo
 \fi
}%
\providecommand \@ifx [1]{%
 \ifx #1\expandafter \@firstoftwo
 \else \expandafter \@secondoftwo
 \fi
}%
\providecommand \natexlab [1]{#1}%
\providecommand \enquote  [1]{``#1''}%
\providecommand \bibnamefont  [1]{#1}%
\providecommand \bibfnamefont [1]{#1}%
\providecommand \citenamefont [1]{#1}%
\providecommand \href@noop [0]{\@secondoftwo}%
\providecommand \href [0]{\begingroup \@sanitize@url \@href}%
\providecommand \@href[1]{\@@startlink{#1}\@@href}%
\providecommand \@@href[1]{\endgroup#1\@@endlink}%
\providecommand \@sanitize@url [0]{\catcode `\\12\catcode `\$12\catcode
  `\&12\catcode `\#12\catcode `\^12\catcode `\_12\catcode `\%12\relax}%
\providecommand \@@startlink[1]{}%
\providecommand \@@endlink[0]{}%
\providecommand \url  [0]{\begingroup\@sanitize@url \@url }%
\providecommand \@url [1]{\endgroup\@href {#1}{\urlprefix }}%
\providecommand \urlprefix  [0]{URL }%
\providecommand \Eprint [0]{\href }%
\providecommand \doibase [0]{http://dx.doi.org/}%
\providecommand \selectlanguage [0]{\@gobble}%
\providecommand \bibinfo  [0]{\@secondoftwo}%
\providecommand \bibfield  [0]{\@secondoftwo}%
\providecommand \translation [1]{[#1]}%
\providecommand \BibitemOpen [0]{}%
\providecommand \bibitemStop [0]{}%
\providecommand \bibitemNoStop [0]{.\EOS\space}%
\providecommand \EOS [0]{\spacefactor3000\relax}%
\providecommand \BibitemShut  [1]{\csname bibitem#1\endcsname}%
\let\auto@bib@innerbib\@empty
%</preamble>
\bibitem [{\citenamefont {Bloch}\ \emph {et~al.}(2008)\citenamefont {Bloch},
  \citenamefont {Dalibard},\ and\ \citenamefont {Zwerger}}]{Bloch08}%
  \BibitemOpen
  \bibfield  {author} {\bibinfo {author} {\bibfnamefont {I.}~\bibnamefont
  {Bloch}}, \bibinfo {author} {\bibfnamefont {J.}~\bibnamefont {Dalibard}}, \
  and\ \bibinfo {author} {\bibfnamefont {W.}~\bibnamefont {Zwerger}},\ }\href
  {\doibase 10.1103/RevModPhys.80.885} {\bibfield  {journal} {\bibinfo
  {journal} {Rev. Mod. Phys.}\ }\textbf {\bibinfo {volume} {80}},\ \bibinfo
  {pages} {885} (\bibinfo {year} {2008})}\BibitemShut {NoStop}%
\bibitem [{\citenamefont {Deng}\ \emph {et~al.}(2010)\citenamefont {Deng},
  \citenamefont {Haug},\ and\ \citenamefont {Yamamoto}}]{Deng10}%
  \BibitemOpen
  \bibfield  {author} {\bibinfo {author} {\bibfnamefont {H.}~\bibnamefont
  {Deng}}, \bibinfo {author} {\bibfnamefont {H.}~\bibnamefont {Haug}}, \ and\
  \bibinfo {author} {\bibfnamefont {Y.}~\bibnamefont {Yamamoto}},\ }\href
  {\doibase 10.1103/RevModPhys.82.1489} {\bibfield  {journal} {\bibinfo
  {journal} {Rev. Mod. Phys.}\ }\textbf {\bibinfo {volume} {82}},\ \bibinfo
  {pages} {1489} (\bibinfo {year} {2010})}\BibitemShut {NoStop}%
\bibitem [{\citenamefont {Lahaye}\ \emph {et~al.}(2009)\citenamefont {Lahaye},
  \citenamefont {Menotti}, \citenamefont {Santos}, \citenamefont {Lewenstein},\
  and\ \citenamefont {Pfau}}]{Lahaye09}%
  \BibitemOpen
  \bibfield  {author} {\bibinfo {author} {\bibfnamefont {T.}~\bibnamefont
  {Lahaye}}, \bibinfo {author} {\bibfnamefont {C.}~\bibnamefont {Menotti}},
  \bibinfo {author} {\bibfnamefont {L.}~\bibnamefont {Santos}}, \bibinfo
  {author} {\bibfnamefont {M.}~\bibnamefont {Lewenstein}}, \ and\ \bibinfo
  {author} {\bibfnamefont {T.}~\bibnamefont {Pfau}},\ }\href@noop {} {\bibfield
   {journal} {\bibinfo  {journal} {Reports on Progress in Physics}\ }\textbf
  {\bibinfo {volume} {72}},\ \bibinfo {pages} {126401} (\bibinfo {year}
  {2009})}\BibitemShut {NoStop}%
\bibitem [{\citenamefont {Pupillo}\ \emph {et~al.}(2010)\citenamefont
  {Pupillo}, \citenamefont {Micheli}, \citenamefont {Boninsegni}, \citenamefont
  {Lesanovsky},\ and\ \citenamefont {Zoller}}]{Pupillo10}%
  \BibitemOpen
  \bibfield  {author} {\bibinfo {author} {\bibfnamefont {G.}~\bibnamefont
  {Pupillo}}, \bibinfo {author} {\bibfnamefont {A.}~\bibnamefont {Micheli}},
  \bibinfo {author} {\bibfnamefont {M.}~\bibnamefont {Boninsegni}}, \bibinfo
  {author} {\bibfnamefont {I.}~\bibnamefont {Lesanovsky}}, \ and\ \bibinfo
  {author} {\bibfnamefont {P.}~\bibnamefont {Zoller}},\ }\href {\doibase
  10.1103/PhysRevLett.104.223002} {\bibfield  {journal} {\bibinfo  {journal}
  {Phys. Rev. Lett.}\ }\textbf {\bibinfo {volume} {104}},\ \bibinfo {pages}
  {223002} (\bibinfo {year} {2010})}\BibitemShut {NoStop}%
\bibitem [{\citenamefont {Laikhtman}\ and\ \citenamefont
  {Rapaport}(2009)}]{Laikhtman09}%
  \BibitemOpen
  \bibfield  {author} {\bibinfo {author} {\bibfnamefont {B.}~\bibnamefont
  {Laikhtman}}\ and\ \bibinfo {author} {\bibfnamefont {R.}~\bibnamefont
  {Rapaport}},\ }\href {\doibase 10.1103/PhysRevB.80.195313} {\bibfield
  {journal} {\bibinfo  {journal} {Phys. Rev. B}\ }\textbf {\bibinfo {volume}
  {80}},\ \bibinfo {pages} {195313} (\bibinfo {year} {2009})}\BibitemShut
  {NoStop}%
\bibitem [{\citenamefont {Astrakharchik}\ \emph {et~al.}(2007)\citenamefont
  {Astrakharchik}, \citenamefont {Boronat}, \citenamefont {Kurbakov},\ and\
  \citenamefont {Lozovik}}]{Astrakharchik07}%
  \BibitemOpen
  \bibfield  {author} {\bibinfo {author} {\bibfnamefont {G.~E.}\ \bibnamefont
  {Astrakharchik}}, \bibinfo {author} {\bibfnamefont {J.}~\bibnamefont
  {Boronat}}, \bibinfo {author} {\bibfnamefont {I.~L.}\ \bibnamefont
  {Kurbakov}}, \ and\ \bibinfo {author} {\bibfnamefont {Y.~E.}\ \bibnamefont
  {Lozovik}},\ }\href {\doibase 10.1103/PhysRevLett.98.060405} {\bibfield
  {journal} {\bibinfo  {journal} {Phys. Rev. Lett.}\ }\textbf {\bibinfo
  {volume} {98}},\ \bibinfo {pages} {060405} (\bibinfo {year}
  {2007})}\BibitemShut {NoStop}%
\bibitem [{\citenamefont {B\"uchler}\ \emph {et~al.}(2007)\citenamefont
  {B\"uchler}, \citenamefont {Demler}, \citenamefont {Lukin}, \citenamefont
  {Micheli}, \citenamefont {Prokof'ev}, \citenamefont {Pupillo},\ and\
  \citenamefont {Zoller}}]{Buchler07}%
  \BibitemOpen
  \bibfield  {author} {\bibinfo {author} {\bibfnamefont {H.~P.}\ \bibnamefont
  {B\"uchler}}, \bibinfo {author} {\bibfnamefont {E.}~\bibnamefont {Demler}},
  \bibinfo {author} {\bibfnamefont {M.}~\bibnamefont {Lukin}}, \bibinfo
  {author} {\bibfnamefont {A.}~\bibnamefont {Micheli}}, \bibinfo {author}
  {\bibfnamefont {N.}~\bibnamefont {Prokof'ev}}, \bibinfo {author}
  {\bibfnamefont {G.}~\bibnamefont {Pupillo}}, \ and\ \bibinfo {author}
  {\bibfnamefont {P.}~\bibnamefont {Zoller}},\ }\href {\doibase
  10.1103/PhysRevLett.98.060404} {\bibfield  {journal} {\bibinfo  {journal}
  {Phys. Rev. Lett.}\ }\textbf {\bibinfo {volume} {98}},\ \bibinfo {pages}
  {060404} (\bibinfo {year} {2007})}\BibitemShut {NoStop}%
\bibitem [{\citenamefont {B\"{o}ning}\ \emph {et~al.}(2011)\citenamefont
  {B\"{o}ning}, \citenamefont {Filinov},\ and\ \citenamefont
  {Bonitz}}]{boening11}%
  \BibitemOpen
  \bibfield  {author} {\bibinfo {author} {\bibfnamefont {J.}~\bibnamefont
  {B\"{o}ning}}, \bibinfo {author} {\bibfnamefont {A.}~\bibnamefont {Filinov}},
  \ and\ \bibinfo {author} {\bibfnamefont {M.}~\bibnamefont {Bonitz}},\ }\href
  {\doibase 10.1103/PhysRevB.84.075130} {\bibfield  {journal} {\bibinfo
  {journal} {Physical Review B}\ }\textbf {\bibinfo {volume} {84}},\ \bibinfo
  {pages} {075130} (\bibinfo {year} {2011})}\BibitemShut {NoStop}%
\bibitem [{\citenamefont {Berman}\ \emph {et~al.}(2012)\citenamefont {Berman},
  \citenamefont {Kezerashvili},\ and\ \citenamefont {Ziegler}}]{Berman12}%
  \BibitemOpen
  \bibfield  {author} {\bibinfo {author} {\bibfnamefont {O.~L.}\ \bibnamefont
  {Berman}}, \bibinfo {author} {\bibfnamefont {R.~Y.}\ \bibnamefont
  {Kezerashvili}}, \ and\ \bibinfo {author} {\bibfnamefont {K.}~\bibnamefont
  {Ziegler}},\ }\href {\doibase 10.1103/PhysRevB.85.035418} {\bibfield
  {journal} {\bibinfo  {journal} {Phys. Rev. B}\ }\textbf {\bibinfo {volume}
  {85}},\ \bibinfo {pages} {035418} (\bibinfo {year} {2012})}\BibitemShut
  {NoStop}%
\bibitem [{\citenamefont {Combescot}\ \emph {et~al.}(2007)\citenamefont
  {Combescot}, \citenamefont {Betbeder-Matibet},\ and\ \citenamefont
  {Combescot}}]{Combescot07}%
  \BibitemOpen
  \bibfield  {author} {\bibinfo {author} {\bibfnamefont {M.}~\bibnamefont
  {Combescot}}, \bibinfo {author} {\bibfnamefont {O.}~\bibnamefont
  {Betbeder-Matibet}}, \ and\ \bibinfo {author} {\bibfnamefont
  {R.}~\bibnamefont {Combescot}},\ }\href {\doibase
  10.1103/PhysRevLett.99.176403} {\bibfield  {journal} {\bibinfo  {journal}
  {Phys. Rev. Lett.}\ }\textbf {\bibinfo {volume} {99}},\ \bibinfo {pages}
  {176403} (\bibinfo {year} {2007})}\BibitemShut {NoStop}%
\bibitem [{\citenamefont {Carr}\ \emph {et~al.}(2009)\citenamefont {Carr},
  \citenamefont {DeMille}, \citenamefont {Krems},\ and\ \citenamefont
  {Ye}}]{Carr09}%
  \BibitemOpen
  \bibfield  {author} {\bibinfo {author} {\bibfnamefont {L.~D.}\ \bibnamefont
  {Carr}}, \bibinfo {author} {\bibfnamefont {D.}~\bibnamefont {DeMille}},
  \bibinfo {author} {\bibfnamefont {R.~V.}\ \bibnamefont {Krems}}, \ and\
  \bibinfo {author} {\bibfnamefont {J.}~\bibnamefont {Ye}},\ }\href@noop {}
  {\bibfield  {journal} {\bibinfo  {journal} {New Journal of Physics}\ }\textbf
  {\bibinfo {volume} {11}},\ \bibinfo {pages} {055049} (\bibinfo {year}
  {2009})}\BibitemShut {NoStop}%
\bibitem [{\citenamefont {Eisenstein}\ and\ \citenamefont
  {{MacDonald}}(2004)}]{eisenstein04}%
  \BibitemOpen
  \bibfield  {author} {\bibinfo {author} {\bibfnamefont {J.~P.}\ \bibnamefont
  {Eisenstein}}\ and\ \bibinfo {author} {\bibfnamefont {A.~H.}\ \bibnamefont
  {{MacDonald}}},\ }\href {\doibase 10.1038/nature03081} {\bibfield  {journal}
  {\bibinfo  {journal} {Nature}\ }\textbf {\bibinfo {volume} {432}},\ \bibinfo
  {pages} {691} (\bibinfo {year} {2004})}\BibitemShut {NoStop}%
\bibitem [{\citenamefont {High}\ \emph {et~al.}(2012)\citenamefont {High},
  \citenamefont {Leonard}, \citenamefont {Hammack}, \citenamefont {Fogler},
  \citenamefont {Butov}, \citenamefont {Kavokin}, \citenamefont {Campman},\
  and\ \citenamefont {Gossard}}]{High12}%
  \BibitemOpen
  \bibfield  {author} {\bibinfo {author} {\bibfnamefont {A.~A.}\ \bibnamefont
  {High}}, \bibinfo {author} {\bibfnamefont {J.~R.}\ \bibnamefont {Leonard}},
  \bibinfo {author} {\bibfnamefont {A.~T.}\ \bibnamefont {Hammack}}, \bibinfo
  {author} {\bibfnamefont {M.~M.}\ \bibnamefont {Fogler}}, \bibinfo {author}
  {\bibfnamefont {L.~V.}\ \bibnamefont {Butov}}, \bibinfo {author}
  {\bibfnamefont {A.~V.}\ \bibnamefont {Kavokin}}, \bibinfo {author}
  {\bibfnamefont {K.~L.}\ \bibnamefont {Campman}}, \ and\ \bibinfo {author}
  {\bibfnamefont {A.~C.}\ \bibnamefont {Gossard}},\ }\href {\doibase
  10.1038/nature10903} {\bibfield  {journal} {\bibinfo  {journal} {Nature}\
  }\textbf {\bibinfo {volume} {483}},\ \bibinfo {pages} {584} (\bibinfo {year}
  {2012})}\BibitemShut {NoStop}%
\bibitem [{\citenamefont {Lee}\ \emph {et~al.}(2009)\citenamefont {Lee},
  \citenamefont {Drummond},\ and\ \citenamefont {Needs}}]{Lee09}%
  \BibitemOpen
  \bibfield  {author} {\bibinfo {author} {\bibfnamefont {R.~M.}\ \bibnamefont
  {Lee}}, \bibinfo {author} {\bibfnamefont {N.~D.}\ \bibnamefont {Drummond}}, \
  and\ \bibinfo {author} {\bibfnamefont {R.~J.}\ \bibnamefont {Needs}},\ }\href
  {\doibase 10.1103/PhysRevB.79.125308} {\bibfield  {journal} {\bibinfo
  {journal} {Phys. Rev. B}\ }\textbf {\bibinfo {volume} {79}},\ \bibinfo
  {pages} {125308} (\bibinfo {year} {2009})}\BibitemShut {NoStop}%
\bibitem [{\citenamefont {Schindler}\ and\ \citenamefont
  {Zimmermann}(2008)}]{Schindler08}%
  \BibitemOpen
  \bibfield  {author} {\bibinfo {author} {\bibfnamefont {C.}~\bibnamefont
  {Schindler}}\ and\ \bibinfo {author} {\bibfnamefont {R.}~\bibnamefont
  {Zimmermann}},\ }\href {\doibase 10.1103/PhysRevB.78.045313} {\bibfield
  {journal} {\bibinfo  {journal} {Phys. Rev. B}\ }\textbf {\bibinfo {volume}
  {78}},\ \bibinfo {pages} {045313} (\bibinfo {year} {2008})}\BibitemShut
  {NoStop}%
\bibitem [{\citenamefont {Stern}\ \emph {et~al.}(2008)\citenamefont {Stern},
  \citenamefont {Garmider}, \citenamefont {Segre}, \citenamefont {Rappaport},
  \citenamefont {Umansky}, \citenamefont {Levinson},\ and\ \citenamefont
  {Bar-Joseph}}]{Stern08B}%
  \BibitemOpen
  \bibfield  {author} {\bibinfo {author} {\bibfnamefont {M.}~\bibnamefont
  {Stern}}, \bibinfo {author} {\bibfnamefont {V.}~\bibnamefont {Garmider}},
  \bibinfo {author} {\bibfnamefont {E.}~\bibnamefont {Segre}}, \bibinfo
  {author} {\bibfnamefont {M.}~\bibnamefont {Rappaport}}, \bibinfo {author}
  {\bibfnamefont {V.}~\bibnamefont {Umansky}}, \bibinfo {author} {\bibfnamefont
  {Y.}~\bibnamefont {Levinson}}, \ and\ \bibinfo {author} {\bibfnamefont
  {I.}~\bibnamefont {Bar-Joseph}},\ }\href {\doibase
  10.1103/PhysRevLett.101.257402} {\bibfield  {journal} {\bibinfo  {journal}
  {Phys. Rev. Lett.}\ }\textbf {\bibinfo {volume} {101}},\ \bibinfo {pages}
  {257402} (\bibinfo {year} {2008})}\BibitemShut {NoStop}%
\bibitem [{\citenamefont {Cohen}\ \emph {et~al.}(2011)\citenamefont {Cohen},
  \citenamefont {Rapaport},\ and\ \citenamefont {Santos}}]{Cohen11}%
  \BibitemOpen
  \bibfield  {author} {\bibinfo {author} {\bibfnamefont {K.}~\bibnamefont
  {Cohen}}, \bibinfo {author} {\bibfnamefont {R.}~\bibnamefont {Rapaport}}, \
  and\ \bibinfo {author} {\bibfnamefont {P.~V.}\ \bibnamefont {Santos}},\
  }\href {\doibase 10.1103/PhysRevLett.106.126402} {\bibfield  {journal}
  {\bibinfo  {journal} {Phys. Rev. Lett.}\ }\textbf {\bibinfo {volume} {106}},\
  \bibinfo {pages} {126402} (\bibinfo {year} {2011})}\BibitemShut {NoStop}%
\bibitem [{\citenamefont {Chen}\ \emph {et~al.}(2006)\citenamefont {Chen},
  \citenamefont {Rapaport}, \citenamefont {Pffeifer}, \citenamefont {West},
  \citenamefont {Platzman}, \citenamefont {Simon}, \citenamefont {V\"or\"os},\
  and\ \citenamefont {Snoke}}]{Chen06}%
  \BibitemOpen
  \bibfield  {author} {\bibinfo {author} {\bibfnamefont {G.}~\bibnamefont
  {Chen}}, \bibinfo {author} {\bibfnamefont {R.}~\bibnamefont {Rapaport}},
  \bibinfo {author} {\bibfnamefont {L.~N.}\ \bibnamefont {Pffeifer}}, \bibinfo
  {author} {\bibfnamefont {K.}~\bibnamefont {West}}, \bibinfo {author}
  {\bibfnamefont {P.~M.}\ \bibnamefont {Platzman}}, \bibinfo {author}
  {\bibfnamefont {S.}~\bibnamefont {Simon}}, \bibinfo {author} {\bibfnamefont
  {Z.}~\bibnamefont {V\"or\"os}}, \ and\ \bibinfo {author} {\bibfnamefont
  {D.}~\bibnamefont {Snoke}},\ }\href {\doibase 10.1103/PhysRevB.74.045309}
  {\bibfield  {journal} {\bibinfo  {journal} {Phys. Rev. B}\ }\textbf {\bibinfo
  {volume} {74}},\ \bibinfo {pages} {045309} (\bibinfo {year}
  {2006})}\BibitemShut {NoStop}%
\bibitem [{\citenamefont {Rapaport}\ \emph {et~al.}(2005)\citenamefont
  {Rapaport}, \citenamefont {Chen}, \citenamefont {Simon}, \citenamefont
  {Mitrofanov}, \citenamefont {Pfeiffer},\ and\ \citenamefont
  {Platzman}}]{Rapaport05}%
  \BibitemOpen
  \bibfield  {author} {\bibinfo {author} {\bibfnamefont {R.}~\bibnamefont
  {Rapaport}}, \bibinfo {author} {\bibfnamefont {G.}~\bibnamefont {Chen}},
  \bibinfo {author} {\bibfnamefont {S.}~\bibnamefont {Simon}}, \bibinfo
  {author} {\bibfnamefont {O.}~\bibnamefont {Mitrofanov}}, \bibinfo {author}
  {\bibfnamefont {L.}~\bibnamefont {Pfeiffer}}, \ and\ \bibinfo {author}
  {\bibfnamefont {P.~M.}\ \bibnamefont {Platzman}},\ }\href {\doibase
  10.1103/PhysRevB.72.075428} {\bibfield  {journal} {\bibinfo  {journal} {Phys.
  Rev. B}\ }\textbf {\bibinfo {volume} {72}},\ \bibinfo {pages} {075428}
  (\bibinfo {year} {2005})}\BibitemShut {NoStop}%
\bibitem [{\citenamefont {Hammack}\ \emph {et~al.}(2006)\citenamefont
  {Hammack}, \citenamefont {Gippius}, \citenamefont {Yang}, \citenamefont
  {Andreev}, \citenamefont {Butov}, \citenamefont {Hanson},\ and\ \citenamefont
  {Gossard}}]{Hammack06}%
  \BibitemOpen
  \bibfield  {author} {\bibinfo {author} {\bibfnamefont {A.~T.}\ \bibnamefont
  {Hammack}}, \bibinfo {author} {\bibfnamefont {N.~A.}\ \bibnamefont
  {Gippius}}, \bibinfo {author} {\bibfnamefont {S.}~\bibnamefont {Yang}},
  \bibinfo {author} {\bibfnamefont {G.~O.}\ \bibnamefont {Andreev}}, \bibinfo
  {author} {\bibfnamefont {L.~V.}\ \bibnamefont {Butov}}, \bibinfo {author}
  {\bibfnamefont {M.}~\bibnamefont {Hanson}}, \ and\ \bibinfo {author}
  {\bibfnamefont {A.~C.}\ \bibnamefont {Gossard}},\ }\href {\doibase
  10.1063/1.2181276} {\bibfield  {journal} {\bibinfo  {journal} {Journal of
  Applied Physics}\ }\textbf {\bibinfo {volume} {99}},\ \bibinfo {eid} {066104}
  (\bibinfo {year} {2006})}\BibitemShut {NoStop}%
\bibitem [{\citenamefont {Schinner}\ \emph {et~al.}(2011)\citenamefont
  {Schinner}, \citenamefont {Schubert}, \citenamefont {Stallhofer},
  \citenamefont {Kotthaus}, \citenamefont {Schuh}, \citenamefont {Rai},
  \citenamefont {Reuter}, \citenamefont {Wieck},\ and\ \citenamefont
  {Govorov}}]{Schinner11}%
  \BibitemOpen
  \bibfield  {author} {\bibinfo {author} {\bibfnamefont {G.~J.}\ \bibnamefont
  {Schinner}}, \bibinfo {author} {\bibfnamefont {E.}~\bibnamefont {Schubert}},
  \bibinfo {author} {\bibfnamefont {M.~P.}\ \bibnamefont {Stallhofer}},
  \bibinfo {author} {\bibfnamefont {J.~P.}\ \bibnamefont {Kotthaus}}, \bibinfo
  {author} {\bibfnamefont {D.}~\bibnamefont {Schuh}}, \bibinfo {author}
  {\bibfnamefont {A.~K.}\ \bibnamefont {Rai}}, \bibinfo {author} {\bibfnamefont
  {D.}~\bibnamefont {Reuter}}, \bibinfo {author} {\bibfnamefont {A.~D.}\
  \bibnamefont {Wieck}}, \ and\ \bibinfo {author} {\bibfnamefont {A.~O.}\
  \bibnamefont {Govorov}},\ }\href {\doibase 10.1103/PhysRevB.83.165308}
  {\bibfield  {journal} {\bibinfo  {journal} {Phys. Rev. B}\ }\textbf {\bibinfo
  {volume} {83}},\ \bibinfo {pages} {165308} (\bibinfo {year}
  {2011})}\BibitemShut {NoStop}%
\bibitem [{\citenamefont {Rapaport}\ \emph {et~al.}(2004)\citenamefont
  {Rapaport}, \citenamefont {Chen}, \citenamefont {Snoke}, \citenamefont
  {Simon}, \citenamefont {Pfeiffer}, \citenamefont {West}, \citenamefont
  {Liu},\ and\ \citenamefont {Denev}}]{Rapaport04}%
  \BibitemOpen
  \bibfield  {author} {\bibinfo {author} {\bibfnamefont {R.}~\bibnamefont
  {Rapaport}}, \bibinfo {author} {\bibfnamefont {G.}~\bibnamefont {Chen}},
  \bibinfo {author} {\bibfnamefont {D.}~\bibnamefont {Snoke}}, \bibinfo
  {author} {\bibfnamefont {S.~H.}\ \bibnamefont {Simon}}, \bibinfo {author}
  {\bibfnamefont {L.}~\bibnamefont {Pfeiffer}}, \bibinfo {author}
  {\bibfnamefont {K.}~\bibnamefont {West}}, \bibinfo {author} {\bibfnamefont
  {Y.}~\bibnamefont {Liu}}, \ and\ \bibinfo {author} {\bibfnamefont
  {S.}~\bibnamefont {Denev}},\ }\href {\doibase 10.1103/PhysRevLett.92.117405}
  {\bibfield  {journal} {\bibinfo  {journal} {Phys. Rev. Lett.}\ }\textbf
  {\bibinfo {volume} {92}},\ \bibinfo {pages} {117405} (\bibinfo {year}
  {2004})}\BibitemShut {NoStop}%
\bibitem [{\citenamefont {Butov}\ \emph {et~al.}(2004)\citenamefont {Butov},
  \citenamefont {Levitov}, \citenamefont {Mintsev}, \citenamefont {Simons},
  \citenamefont {Gossard},\ and\ \citenamefont {Chemla}}]{Butov04prl}%
  \BibitemOpen
  \bibfield  {author} {\bibinfo {author} {\bibfnamefont {L.~V.}\ \bibnamefont
  {Butov}}, \bibinfo {author} {\bibfnamefont {L.~S.}\ \bibnamefont {Levitov}},
  \bibinfo {author} {\bibfnamefont {A.~V.}\ \bibnamefont {Mintsev}}, \bibinfo
  {author} {\bibfnamefont {B.~D.}\ \bibnamefont {Simons}}, \bibinfo {author}
  {\bibfnamefont {A.~C.}\ \bibnamefont {Gossard}}, \ and\ \bibinfo {author}
  {\bibfnamefont {D.~S.}\ \bibnamefont {Chemla}},\ }\href {\doibase
  10.1103/PhysRevLett.92.117404} {\bibfield  {journal} {\bibinfo  {journal}
  {Phys. Rev. Lett.}\ }\textbf {\bibinfo {volume} {92}},\ \bibinfo {pages}
  {117404} (\bibinfo {year} {2004})}\BibitemShut {NoStop}%
\bibitem [{\citenamefont {Sivalertporn}\ \emph {et~al.}(2012)\citenamefont
  {Sivalertporn}, \citenamefont {Mouchliadis}, \citenamefont {Ivanov},
  \citenamefont {Philp},\ and\ \citenamefont {Muljarov}}]{Sivalertporn12}%
  \BibitemOpen
  \bibfield  {author} {\bibinfo {author} {\bibfnamefont {K.}~\bibnamefont
  {Sivalertporn}}, \bibinfo {author} {\bibfnamefont {L.}~\bibnamefont
  {Mouchliadis}}, \bibinfo {author} {\bibfnamefont {A.~L.}\ \bibnamefont
  {Ivanov}}, \bibinfo {author} {\bibfnamefont {R.}~\bibnamefont {Philp}}, \
  and\ \bibinfo {author} {\bibfnamefont {E.~A.}\ \bibnamefont {Muljarov}},\
  }\href {\doibase 10.1103/PhysRevB.85.045207} {\bibfield  {journal} {\bibinfo
  {journal} {Phys. Rev. B}\ }\textbf {\bibinfo {volume} {85}},\ \bibinfo
  {pages} {045207} (\bibinfo {year} {2012})}\BibitemShut {NoStop}%
\bibitem [{\citenamefont {Maialle}\ \emph {et~al.}(1993)\citenamefont
  {Maialle}, \citenamefont {de~Andrada~e Silva},\ and\ \citenamefont
  {Sham}}]{Maialle93}%
  \BibitemOpen
  \bibfield  {author} {\bibinfo {author} {\bibfnamefont {M.~Z.}\ \bibnamefont
  {Maialle}}, \bibinfo {author} {\bibfnamefont {E.~A.}\ \bibnamefont
  {de~Andrada~e Silva}}, \ and\ \bibinfo {author} {\bibfnamefont {L.~J.}\
  \bibnamefont {Sham}},\ }\href {\doibase 10.1103/PhysRevB.47.15776} {\bibfield
   {journal} {\bibinfo  {journal} {Phys. Rev. B}\ }\textbf {\bibinfo {volume}
  {47}},\ \bibinfo {pages} {15776} (\bibinfo {year} {1993})}\BibitemShut
  {NoStop}%
\bibitem [{\citenamefont {Piermarocchi}\ \emph {et~al.}(1997)\citenamefont
  {Piermarocchi}, \citenamefont {Tassone}, \citenamefont {Savona},
  \citenamefont {Quattropani},\ and\ \citenamefont
  {Schwendimann}}]{Piermarocchi97}%
  \BibitemOpen
  \bibfield  {author} {\bibinfo {author} {\bibfnamefont {C.}~\bibnamefont
  {Piermarocchi}}, \bibinfo {author} {\bibfnamefont {F.}~\bibnamefont
  {Tassone}}, \bibinfo {author} {\bibfnamefont {V.}~\bibnamefont {Savona}},
  \bibinfo {author} {\bibfnamefont {A.}~\bibnamefont {Quattropani}}, \ and\
  \bibinfo {author} {\bibfnamefont {P.}~\bibnamefont {Schwendimann}},\ }\href
  {\doibase 10.1103/PhysRevB.55.1333} {\bibfield  {journal} {\bibinfo
  {journal} {Phys. Rev. B}\ }\textbf {\bibinfo {volume} {55}},\ \bibinfo
  {pages} {1333} (\bibinfo {year} {1997})}\BibitemShut {NoStop}%
\bibitem [{\citenamefont {Butov}\ \emph {et~al.}(1999)\citenamefont {Butov},
  \citenamefont {Shashkin}, \citenamefont {Dolgopolov}, \citenamefont
  {Campman},\ and\ \citenamefont {Gossard}}]{Butov99}%
  \BibitemOpen
  \bibfield  {author} {\bibinfo {author} {\bibfnamefont {L.~V.}\ \bibnamefont
  {Butov}}, \bibinfo {author} {\bibfnamefont {A.~A.}\ \bibnamefont {Shashkin}},
  \bibinfo {author} {\bibfnamefont {V.~T.}\ \bibnamefont {Dolgopolov}},
  \bibinfo {author} {\bibfnamefont {K.~L.}\ \bibnamefont {Campman}}, \ and\
  \bibinfo {author} {\bibfnamefont {A.~C.}\ \bibnamefont {Gossard}},\ }\href
  {\doibase 10.1103/PhysRevB.60.8753} {\bibfield  {journal} {\bibinfo
  {journal} {Phys. Rev. B}\ }\textbf {\bibinfo {volume} {60}},\ \bibinfo
  {pages} {8753} (\bibinfo {year} {1999})}\BibitemShut {NoStop}%
\bibitem [{\citenamefont {Leonard}\ \emph {et~al.}(2009)\citenamefont
  {Leonard}, \citenamefont {Kuznetsova}, \citenamefont {Yang}, \citenamefont
  {Butov}, \citenamefont {Ostatnicky´}, \citenamefont {Kavokin},\ and\
  \citenamefont {Gossard}}]{Leonard10}%
  \BibitemOpen
  \bibfield  {author} {\bibinfo {author} {\bibfnamefont {J.~R.}\ \bibnamefont
  {Leonard}}, \bibinfo {author} {\bibfnamefont {Y.~Y.}\ \bibnamefont
  {Kuznetsova}}, \bibinfo {author} {\bibfnamefont {S.}~\bibnamefont {Yang}},
  \bibinfo {author} {\bibfnamefont {L.~V.}\ \bibnamefont {Butov}}, \bibinfo
  {author} {\bibfnamefont {T.}~\bibnamefont {Ostatnicky´}}, \bibinfo {author}
  {\bibfnamefont {A.}~\bibnamefont {Kavokin}}, \ and\ \bibinfo {author}
  {\bibfnamefont {A.~C.}\ \bibnamefont {Gossard}},\ }\href {\doibase
  10.1021/nl9024227} {\bibfield  {journal} {\bibinfo  {journal} {Nano Letters}\
  }\textbf {\bibinfo {volume} {9}},\ \bibinfo {pages} {4204} (\bibinfo {year}
  {2009})}\BibitemShut {NoStop}%
\end{thebibliography}

\begin{thebibliography}{4}%
\makeatletter
\providecommand \@ifxundefined [1]{%
 \@ifx{#1\undefined}
}%
\providecommand \@ifnum [1]{%
 \ifnum #1\expandafter \@firstoftwo
 \else \expandafter \@secondoftwo
 \fi
}%
\providecommand \@ifx [1]{%
 \ifx #1\expandafter \@firstoftwo
 \else \expandafter \@secondoftwo
 \fi
}%
\providecommand \natexlab [1]{#1}%
\providecommand \enquote  [1]{``#1''}%
\providecommand \bibnamefont  [1]{#1}%
\providecommand \bibfnamefont [1]{#1}%
\providecommand \citenamefont [1]{#1}%
\providecommand \href@noop [0]{\@secondoftwo}%
\providecommand \href [0]{\begingroup \@sanitize@url \@href}%
\providecommand \@href[1]{\@@startlink{#1}\@@href}%
\providecommand \@@href[1]{\endgroup#1\@@endlink}%
\providecommand \@sanitize@url [0]{\catcode `\\12\catcode `\$12\catcode
  `\&12\catcode `\#12\catcode `\^12\catcode `\_12\catcode `\%12\relax}%
\providecommand \@@startlink[1]{}%
\providecommand \@@endlink[0]{}%
\providecommand \url  [0]{\begingroup\@sanitize@url \@url }%
\providecommand \@url [1]{\endgroup\@href {#1}{\urlprefix }}%
\providecommand \urlprefix  [0]{URL }%
\providecommand \Eprint [0]{\href }%
\providecommand \doibase [0]{http://dx.doi.org/}%
\providecommand \selectlanguage [0]{\@gobble}%
\providecommand \bibinfo  [0]{\@secondoftwo}%
\providecommand \bibfield  [0]{\@secondoftwo}%
\providecommand \translation [1]{[#1]}%
\providecommand \BibitemOpen [0]{}%
\providecommand \bibitemStop [0]{}%
\providecommand \bibitemNoStop [0]{.\EOS\space}%
\providecommand \EOS [0]{\spacefactor3000\relax}%
\providecommand \BibitemShut  [1]{\csname bibitem#1\endcsname}%
\let\auto@bib@innerbib\@empty
%</preamble>
\bibitem [{\citenamefont {Hagn}\ \emph {et~al.}(1995)\citenamefont {Hagn},
  \citenamefont {Zrenner}, \citenamefont {Böhm},\ and\ \citenamefont
  {Weimann}}]{Hagn95}%
  \BibitemOpen
  \bibfield  {author} {\bibinfo {author} {\bibfnamefont {M.}~\bibnamefont
  {Hagn}}, \bibinfo {author} {\bibfnamefont {A.}~\bibnamefont {Zrenner}},
  \bibinfo {author} {\bibfnamefont {G.}~\bibnamefont {Böhm}}, \ and\ \bibinfo
  {author} {\bibfnamefont {G.}~\bibnamefont {Weimann}},\ }\href {\doibase
  doi:10.1063/1.114677} {\bibfield  {journal} {\bibinfo  {journal} {Applied
  Physics Letters}\ }\textbf {\bibinfo {volume} {67}},\ \bibinfo {pages} {232}
  (\bibinfo {year} {1995})}\BibitemShut {NoStop}%
\bibitem [{\citenamefont {Rapaport}\ \emph {et~al.}(2005)\citenamefont
  {Rapaport}, \citenamefont {Chen}, \citenamefont {Simon}, \citenamefont
  {Mitrofanov}, \citenamefont {Pfeiffer},\ and\ \citenamefont
  {Platzman}}]{Rapaport05}%
  \BibitemOpen
  \bibfield  {author} {\bibinfo {author} {\bibfnamefont {R.}~\bibnamefont
  {Rapaport}}, \bibinfo {author} {\bibfnamefont {G.}~\bibnamefont {Chen}},
  \bibinfo {author} {\bibfnamefont {S.}~\bibnamefont {Simon}}, \bibinfo
  {author} {\bibfnamefont {O.}~\bibnamefont {Mitrofanov}}, \bibinfo {author}
  {\bibfnamefont {L.}~\bibnamefont {Pfeiffer}}, \ and\ \bibinfo {author}
  {\bibfnamefont {P.~M.}\ \bibnamefont {Platzman}},\ }\href {\doibase
  10.1103/PhysRevB.72.075428} {\bibfield  {journal} {\bibinfo  {journal} {Phys.
  Rev. B}\ }\textbf {\bibinfo {volume} {72}},\ \bibinfo {pages} {075428}
  (\bibinfo {year} {2005})}\BibitemShut {NoStop}%
\bibitem [{\citenamefont {Hammack}\ \emph {et~al.}(2006)\citenamefont
  {Hammack}, \citenamefont {Gippius}, \citenamefont {Yang}, \citenamefont
  {Andreev}, \citenamefont {Butov}, \citenamefont {Hanson},\ and\ \citenamefont
  {Gossard}}]{Hammack06}%
  \BibitemOpen
  \bibfield  {author} {\bibinfo {author} {\bibfnamefont {A.~T.}\ \bibnamefont
  {Hammack}}, \bibinfo {author} {\bibfnamefont {N.~A.}\ \bibnamefont
  {Gippius}}, \bibinfo {author} {\bibfnamefont {S.}~\bibnamefont {Yang}},
  \bibinfo {author} {\bibfnamefont {G.~O.}\ \bibnamefont {Andreev}}, \bibinfo
  {author} {\bibfnamefont {L.~V.}\ \bibnamefont {Butov}}, \bibinfo {author}
  {\bibfnamefont {M.}~\bibnamefont {Hanson}}, \ and\ \bibinfo {author}
  {\bibfnamefont {A.~C.}\ \bibnamefont {Gossard}},\ }\href {\doibase
  10.1063/1.2181276} {\bibfield  {journal} {\bibinfo  {journal} {Journal of
  Applied Physics}\ }\textbf {\bibinfo {volume} {99}},\ \bibinfo {eid} {066104}
  (\bibinfo {year} {2006})}\BibitemShut {NoStop}%
\bibitem [{\citenamefont {Kowalik-Seidl}\ \emph {et~al.}(2012)\citenamefont
  {Kowalik-Seidl}, \citenamefont {Vögele}, \citenamefont {Rimpfl},
  \citenamefont {Schinner}, \citenamefont {Schuh}, \citenamefont {Wegscheider},
  \citenamefont {Holleitner},\ and\ \citenamefont
  {Kotthaus}}]{Kowalik-Seidl12}%
  \BibitemOpen
  \bibfield  {author} {\bibinfo {author} {\bibfnamefont {K.}~\bibnamefont
  {Kowalik-Seidl}}, \bibinfo {author} {\bibfnamefont {X.~P.}\ \bibnamefont
  {Vögele}}, \bibinfo {author} {\bibfnamefont {B.~N.}\ \bibnamefont {Rimpfl}},
  \bibinfo {author} {\bibfnamefont {G.~J.}\ \bibnamefont {Schinner}}, \bibinfo
  {author} {\bibfnamefont {D.}~\bibnamefont {Schuh}}, \bibinfo {author}
  {\bibfnamefont {W.}~\bibnamefont {Wegscheider}}, \bibinfo {author}
  {\bibfnamefont {A.~W.}\ \bibnamefont {Holleitner}}, \ and\ \bibinfo {author}
  {\bibfnamefont {J.~P.}\ \bibnamefont {Kotthaus}},\ }\href {\doibase
  10.1021/nl203613k} {\bibfield  {journal} {\bibinfo  {journal} {Nano Letters}\
  }\textbf {\bibinfo {volume} {12}},\ \bibinfo {pages} {326} (\bibinfo {year}
  {2012})}\BibitemShut {NoStop}%
\end{thebibliography}
\end{document}